\documentclass[reprint,aps,prd]{revtex4-2}
\usepackage{amsmath,amssymb,amsthm,graphicx}
\usepackage[mathscr]{eucal}
\usepackage[colorlinks,linkcolor=blue,anchorcolor=blue,citecolor=green]{hyperref}
\usepackage[dvipsnames,usenames]{color}

\usepackage{physics}
\usepackage{lipsum}
\usepackage{epsf,epsfig,graphics,soul}
\usepackage{verbatim,color,ulem,booktabs}
\usepackage{orcidlink}
\newcommand{\beq}{\begin{equation}}
\newcommand{\eeq}{\end{equation}}

\begin{document}
\title{Regge poles  of analogous rotating black holes in binary
Bose-Einstein condensates: The gapped excitations}
\date{\today}
\author{Wei-Can Syu \orcidlink{0000-0001-8359-4219}}
\email{syuweican@gmail.com}
\affiliation{Center of General Education, Wenzao Ursuline University of Languages, Kaohsiung 780793, Taiwan, R.O.C.}
\author{Tien Hsieh \orcidlink{0000-0001-7199-1241}}
\affiliation{Department of
	Physics, National Dong-Hwa University, Hualien 974301, Taiwan, R.O.C.}

\author{Da-Shin Lee \orcidlink{0000-0003-3187-8863}}
\email{dslee@gms.ndhu.edu.tw}
\affiliation{Department of
	Physics, National Dong-Hwa University, Hualien 974301, Taiwan, R.O.C.}

\begin{abstract}
In this paper, we study the spectrum of the Regge poles (RPs), which are the counterparts of quasinormal modes,  in a draining bathtub vortex within a two-component Bose-Einstein condensate (BEC) system.
We study the gapped excitations of the condensate with the spatially dependent energy gap term using a spatially tunable Rabi coupling, which will be treated as a perturbation. This model serves as an analog of a rotating black hole surrounded by an environmental mass shell. We first compute the semiclassical scattering amplitude with the spatially independent mass effect due to the orbital interference. In the case of the mass-shell,  bifurcation of the spectrum is observed, resulting in the destabilization of the RPs. We also study the migration of RPs by shifting the bump position.
Our results show that the RPs of the corotating modes exhibit greater stability than those of the counterrotating  modes. Large migration and overtaking jumps of the overtone (fundamental RP) leave an imprint on the scattering amplitude at small (large) scattering angles.   This can be observed in the scattering interference pattern in experiments.
\end{abstract}

\maketitle
\newpage
\section{INTRODUCTION}\label{sec_introduction}
In general relativity (GR)  linear gravitational perturbations  of  black hole
spacetimes can be probed through their damped oscillatory
behavior, giving rise to the emission of gravitational waves during the ringdown phase  of a binary black hole merger.
They are of particular importance today, because of their relevance to gravitational wave astronomy. The damped resonances are manifest
as poles in the scattering matrix, occurring at
complex frequencies and complex angular momenta.
The corresponding modes are known as quasinormal modes
(QNM) and Regge pole (RP), respectively \cite{Berti2009,Konoplya2011,Folacci2019}.
From the perspective of RPs, they will provide a framework to establish a link between spectral instabilities and scattering quantities.
Nevertheless, black holes are typically not isolated, but are surrounded by matter such as accretion disks and dark matter. Therefore, the QNM spectra that we observe from gravitational waves should be modified by interactions with their surroundings \cite{Barausse2014,Cardoso2020,McGee2020,Cardoso2013,Cardoso2013b,Lingetti2022}.

 Recently studies in analog gravity  have  focused on mimicking  the geometry of a spinning  black hole in $2+1$ dimensions using a draining bathtub (DBT) vortex flow \cite{Visser1998,Richartz2015,Torres2017,Torres2020,Berti2023,Patrick2022}. Apart from  a theoretical consideration of  massless scalar fields in the analog  black hole spacetime,
 the massive scalar fields  has also been put forward \cite{Fischer2004,Liberati2006,Visser2005,Syu2022,Syu2023,Syu2019,Syu2024}. In particular, in Ref.~\cite{Syu2024}, we consider two-component Bose-Einstein condensates (BECs) and introduce the BEC vortex in $2+1$ dimensions.
 The advent of  experimental studies on tunable binary BECs in Refs.~\cite{Kim2020,Cominotti2022,Hamner2011,Hamner2013} makes it possible  to observe these two types of excitations.
 Our primary focus is on gapped excitations, in which the  energy gap acts as a mass term for an analog of relativistic scalar fields.
 We study superradiant instabilities resulting from the quasibound states due to positive mass squared and the tachyonic instabilities arising from negative mass squared in both the frequency and time domains.
 Similarly, the dynamical instability in analog systems has been studied \cite{Finazzi2010, Coutant2010, Coutant2016}.
In this work, we consider RPs and explore potential spectral instabilities of gapped excitations in the presence of a rotating vortex in a DBT background.   Building on the work of Refs.~\cite{Dolan2012, Dolan2013b}, we extend it to incorporate spatially shell-like perturbations, giving rise to the mass shell of an analogous scalar field.

With the spatially dependent perturbations, the spectral instabilities are found when bifurcations of the complex angular momentum appears  in the spherically symmetric spacetime \cite{Torres2023, Torres2023b}.
One of our main focuses here is the potential destabilization behavior of RPs in an analogous rotating black hole spacetime.
 Additionally, we apply the idea of Ref.~\cite{Cardoso2022}  to examine the migration of RPs in response to changes in the location of the mass shell.
 This allows us to investigate the behavior of the bifurcation of RPs, which may undergo either large migration or the so-called overtaking jumps.  The instability behavior in terms of time-independent scattering can also be reconstructed using the  complex angular momentum (CAM) method.

The layout of the paper is as follows. In Sec.~\ref{secii},
  we introduce the two-component BEC system, adapted from \cite{Syu2024}, and the time-independent scattering theory.
In Sec.~\ref{seciii}, the scattering amplitude is computed using a  semiclassical approximation for high-frequency scattering, This provides an alternative interpretation of the scattering amplitude in terms of orbiting.
In Sec.~\ref{seciv}, the RPs of the fundamental mode and overtones are studied, revealing  bifurcation of the RP spectrum when taking account of the bump perturbations.
Additionally, the migration trajectory and how changing the bump location affects it will be explored.
Section~\ref{secv}  is devoted to the numerical study.
In Sec.~\ref{secvi}, the conclusions are drawn.
Appendix~\ref{appA} reviews the continued fraction method (CFM), and
Appendix~\ref{appB} provides more detailed derivations to achieve at (\ref{lambda}).

\section{set-up and Time-independent scattering theory }\label{secii}
Here, we provide a minireview of the DBT model in two-component BECs and summarize the key equations. See \cite{Syu2024} for more details. The binary Bose-Einstein condensates system we consider is for identical atoms with distinct internal hyperfine states, where the coupling constants can be tuned by introducing Feshbach resonances~\cite{Myatt1997,Hall1998,Papp2008,Tojo2010}.
In addition, transitions between these two states are allowed and governed by the Rabi frequency $\Omega$, where $\Omega\ge0$. With the unit $\hbar=1$ throughout this paper, the coupled time-dependent equations of motion in $2+1$ dimensions are expressed by
\begin{align}\label{GP}
		 i\partial_t\hat{\Psi}_i=\left[-\frac{1}{2m_a}\nabla^2+V_i+g_{ii}\hat{\Psi}_i^\dagger\hat{\Psi}_i+g_{ij}\hat{\Psi}_j^\dagger\hat{\Psi}_j\right]\hat{\Psi}_i-\frac{\Omega}{2}\hat{\Psi}_j\,
\end{align}
with $i\neq j$ and $i,j=1,2$,  where $m_a$  is the atomic mass, and $V_i$ are the external potentials corresponding  to the hyperfine states $i$.
The coupling constants $g_{ii}$ and $g_{ij}$ represent the strength of the interaction between atoms in the same hyperfine state and between atoms in different hyperfine states, respectively.

 The condensate wave functions can be obtained from the expectation values of the field operator $\langle\hat{\Psi}_i \rangle$  given by,
$ \langle \hat{\Psi}_i\rangle(\mathbf{ r},t)=\sqrt{\rho_{i}(\mathbf{ r},t)}\, e^{ i\theta_{i}(\mathbf{ r},t)-i \mu t}$
 with the chemical potential $\mu$.
 The condensate flow velocities are  $ \mathbf{v}_i={\grad} \theta_{i} /m_a$.
The perturbations around the stationary wave functions can be defined through
$ \hat{\Psi}_i=\langle\hat{\Psi}_i \rangle(1+\hat{\phi}_i) \, ,$
{where 	the fluctuation fields can be written   in terms of the density and the phase  as $
 		\hat{\phi}_i=\delta{\hat{n}_i}+i\delta\hat{\theta}_i=\frac{\delta\hat{\rho}_i}{2\rho_{i}}+i\delta\hat{\theta}_i\,.$
 		
According to \cite{Visser2005}, the equations can be decoupled for the general spatially dependent condensate wave functions and the coupling strengths by choosing the time-independent background solutions as $\rho_{1}=\rho_2=\rho$, $\theta_{1}=\theta_2=\theta$, and
$g_{11}=g_{22}=g$.
We also consider a miscible state of the background condensates with the scattering parameters of $g_{12}<g$ .
 Further linear combination between the states
$ \sqrt{2}\delta \hat n_{d/s}=\delta \hat{n}_1\pm\delta \hat{n}_2$ and
$ \sqrt{2}\delta \hat {\theta}_{d/s}=\delta \hat{\theta}_1 \pm\delta \hat{\theta}_2$ giving the (gapless) density and (gapped) spin modes
yields a single equation for $\delta\hat\theta_s$
in the form of  the Klein-Gordon  equation for a massive scalar field,
\begin{align}
\frac{1}{\sqrt{-\mathbf{g}}}\partial_\mu\left(\sqrt{-\mathbf{g}}\,\mathbf{g}^{\mu\nu}\partial_\nu \delta\hat\theta_s\right)-m_\text{eff}^2\delta\hat\theta_s=0,
\label{KGE}
\end{align}
where the acoustic metric $\mathbf{g}^{\mu\nu}$ depends on the background density $\rho$ and  phase $\theta$ with the spatially dependent sound speed given by  $c^2=[(g-g_{12})\rho+\Omega]/m_a\simeq (g-g_{12})\rho/m_a$, and  the effective mass  $m^2_\text{eff}= 2m_a^2 c\,\Omega/\rho $.  In this work, we consider small perturbations given by the small $\Omega$, which mimic the environmental effects to the dynamics of classical perturbations around black hole spacetime.

Following \cite{Patrick2022}, let us build up the analogs geometry of a rotating black hole by introducing a draining vortex with the velocity flow
\begin{align}
	\mathbf{v}=\frac{1}{m_a}{\grad}\theta=\frac{-d \mathbf{e}_r+\ell \mathbf{e}_\phi}{m_a r}=v_r\mathbf{e}_r+v_\phi\mathbf{e}_\phi,
\end{align}
where the winding number $\ell$ is taken as an integer for quantum vortices and $d>0$ for the draining flow.
In the Thomas-Fermi approximations,  we consider the region  far from the core of the vortex with the radius $r \gg R_\text{TF}= \sqrt{\frac{d^2+\ell^2}{2m_a \rho_\infty(g+g_{12})}}$,
where  the density $\rho$ can be treated as a constant \cite{Patrick2022}, $\rho \simeq \rho_\infty$ given by
$
\rho_\infty=\frac{1}{g+g_{12}}\left(\mu-V_0 \pm\frac{\Omega}{2}\right) \simeq \frac{1}{g+g_{12}}\left(\mu-V_0 \right)$ for an uniform external potential $V_0$ and by ignoring the small $\Omega$, leading to a constant sound speed $c^2$.
Under the coordinate transformations
\begin{subequations}
\begin{align}
	d\tilde{t}& =dt-(c^2/v_r^2-1)^{-1}dr\, , \\
	d\tilde{\phi}& =d\phi-\frac{v_rv_\phi}{r(c^2-v_r^2)}dr,
\end{align}
\end{subequations}
the metric becomes
\begin{align} ds^2=&\,\tilde g_{\mu\nu}^\text{vortex}d \tilde x^\mu  d \tilde x^\nu \nonumber\\ \propto&-\left(1-\frac{r_E^2}{r^2}\right)d\tilde{t}^2+\left(1-\frac{r^2_H}{r^2}\right)^{-1}dr^2\nonumber\\
	&-\frac{2\ell}{m_a}\, d\tilde{\phi}d\tilde{t}+r^2d\tilde{\phi} ^2\, \label{metric}
\end{align}
with $\tilde x=(\tilde t, r, \tilde \phi)$, where the acoustic horizon and the ergosphere are located at $r_\text{H}=d/m_ac$ and $r_\text{E}=\sqrt{d^2+\ell^2}/m_ac$, respectively.
The notation can be simplified by defining dimensionless variables:
$r/r_H\rightarrow r$, $c\,\tilde {t}/r_H\rightarrow  {t}$ and $r_H\Omega/c\rightarrow\Omega$. For convenience, we also relabel $\tilde{\phi}$ as $\phi$.

\begin{figure}[t]
	\includegraphics[width=0.8\linewidth]{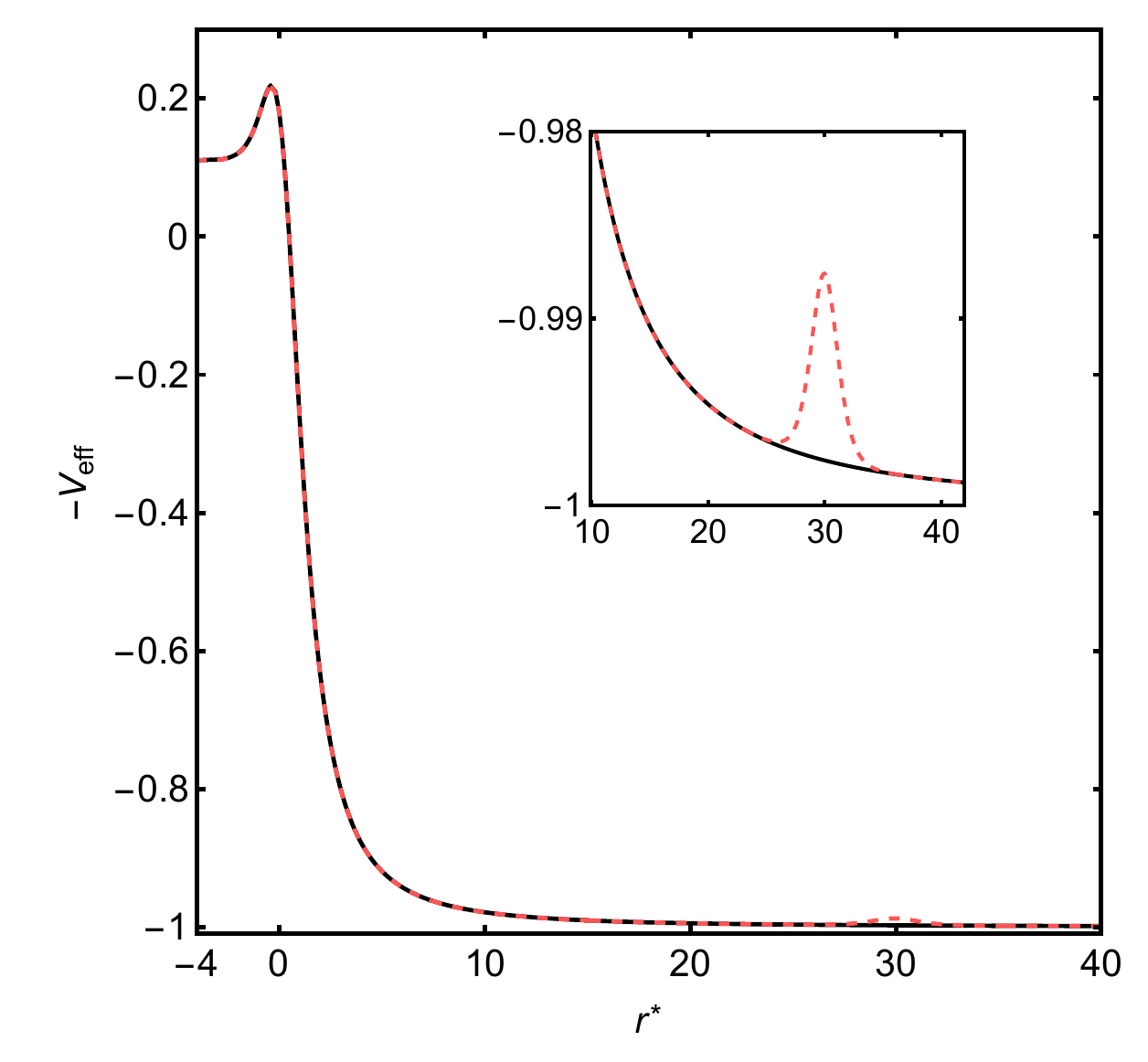}
	\caption{The effective potential as a function of the tortoise coordinate $r^\ast$ without the perturbation (black) and with the perturbation (red dashed) at $r_0=30$.  The parameters are set to $\omega=1.0$ and $\ell=1$. The inset plot zooms in on the effective potential near the bump. Other parameters are fixed as $\alpha=3/2$, $\epsilon=10^{-2}$.}
	\label{fig_veff}
\end{figure}
Using the field of phase fluctuation  of the form
\begin{equation}
	\delta\theta_s({t},r,{\phi})=\frac{H_{\omega,m}(r)}{\sqrt{r}}\, e^{ -i\omega t+i m{\phi}} \, ,
\end{equation}
 the radial equation in the tortoise coordinate
$
	r^\ast=r+\frac{1}{2}\log{\Big\vert\frac{r-1}{r+1}\Big\vert}
$ becomes  the time-independent Schrodinger-like equation
 \begin{align}
\left[\partial_{r^\ast}^2+V_\text{eff}(r)\right]H_{\omega,m}(r)=0,
\label{WEQ}
\end{align}
 where
\begin{align} V_\text{eff}(r)=&\left({\omega-\frac{m\ell}{r^2}}\right)^2-\left(\frac{r^2-1}{r^2}\right)\nonumber\\
	&\times\left[\frac{5}{4r^4}+\frac{1}{r^2}\left(m^2-\frac{1}{4}\right)+\mu^2_s(r)\right]\label{Veff2}
\end{align}
with $\mu^2_s(r)= 2 d \Omega(r)$.
The effective mass squared term can now be parametrized in an analytically treatable form as
\begin{align}
	\mu^2_s (r)=2d\Omega(r)= \epsilon\,\text{sech}^2[\alpha(r-r_0)],\label{mass}
\end{align}
where $\epsilon=2d\Omega_0$,
 mimicking a mass shell surrounding the black hole \cite{Cardoso2013,Cardoso2013b,Cheung2022,Courty2023}.
 The profile is controlled by its typical distance from the black hole horizon and the width of the distribution. Throughout this paper, we fix the parameters as $\epsilon=10^{-2}$, and $\alpha=3/2$. The typical $V_\text{eff}(r)$ is shown in Fig.~\ref{fig_veff}.

The theoretical studies of a planar wave scattered by a draining vortex has been intensely studied in Refs.~\cite{Dolan2012,Dolan2013b}.
 Here we numerically investigate  the wave equation \eqref{WEQ}.  The solution  obeys the boundary conditions of the pure incoming wave at the horizon for the modes in the hydrodynamic limit with $k \xi \le 1$ where the healing length $ \xi =1/m_a c$  and the incoming/outgoing waves at infinity. It is given by
\begin{align}
	H_{\omega,m}(r^\ast)\sim \begin{cases}
		e^{-i(\omega-m\ell) r^\ast}, &  (r^\ast\rightarrow -\infty),\\
		A^\text{out}e^{i\omega r^\ast}+A^\text{in}e^{-i\omega r^\ast}, &  (r^\ast\rightarrow \infty).
	\end{cases}
\label{ingoing}
\end{align}
The scattering amplitude is given by
\begin{align}
	f_\omega(\phi)=\sqrt{\frac{1}{2i\pi\omega}}\sum_{m=-\infty}^{\infty}\left(e^{2i\delta_m}-1\right)e^{im\phi}
	\label{pws}
\end{align}
with the phase shift defined as
\begin{align}
	e^{2i\delta_m}=i(-1)^mA^\text{out}/A^\text{in}\, .
\end{align}
The differential cross section is obtained from the scattering amplitude as
\begin{align}\label{d_cross_sec}
	\frac{d\sigma}{d\phi}=\vert f_\omega(\phi)\vert^2\, .
\end{align}

For a real-valued frequency $\omega$ and  with the pure outgoing wave $e^{i\omega r^\ast}$ at infinity, namely $A^{\rm in}=0$, the resonances occur with the complex value  $m={\rm Re}\,  m+i{\rm Im} \, m$ for the  RPs, with which to compute  the time-independent scattering amplitude.
\section{differential scattering cross section in semiclassical approximations}\label{seciii}
\begin{figure}[t]
\centering
\includegraphics[width=0.9\linewidth]{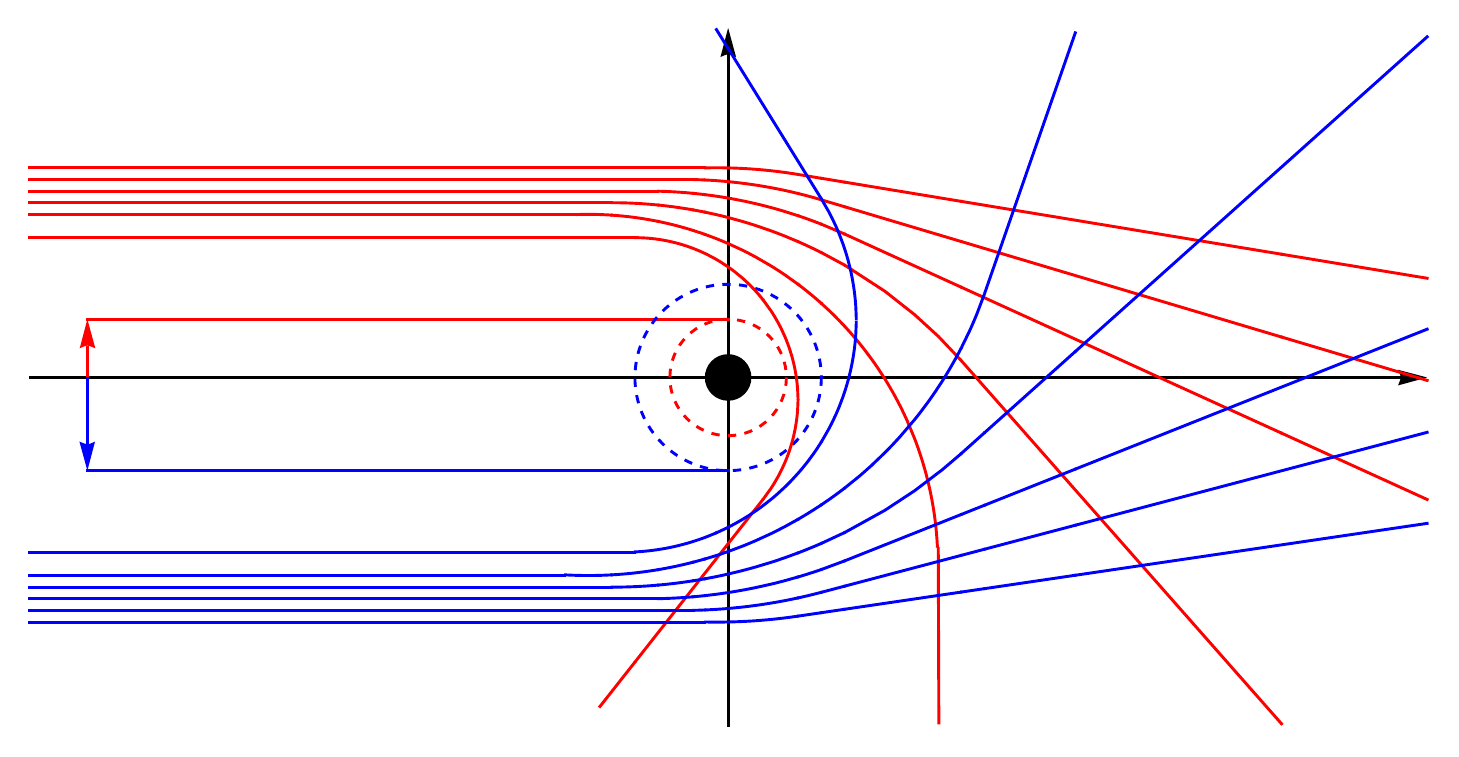}
    \caption{Schematic diagram of the particle scattering with the interference. The spin of the black hole is clockwise.  The red solid line represents the corotating orbits, while the blue solid line corresponds to the counterrotating  orbits. }
\label{Schematic}
\end{figure}
In quantum mechanics, the wave properties will lead to the fact that the differential cross section is  the square modulus of the scattering amplitude, which is computed as the sum of  wave functions that scatter into  a particular  angle.
Before computing the full quantum mechanical scattering amplitude, we begin by studying it in the semiclassical approximation where the geodesic dynamics in the effective rotating black hole spacetime as well as  the interference effects between them are considered \cite{Ford1959,Nussenzveig1992}.  This semiclassical differential  scattering amplitude will show the characteristic properties of the full quantum mechanical scattering amplitude, which is suited to the relatively high frequency scattering.

Below, we will follow the semiclassical analysis in Ref.~\cite{Dolan2013b} but extend it by including the perturbation to the null geodesics arising from an effective small spatially independent mass term, which is theoretically treatable.
We consider  high energy scattering, which the energy is of order of the inverse of the characteristic length scale of the analog rotating spacetime, the ergosphere radius,   such that $E\propto 1/r_E \gg \sqrt\epsilon$. In an asymptotic regime, the energy can be safely approximated by   $ E^2=\omega^2+\epsilon^2 \approx \omega^2$.

However, the geodesics equation, governing the orbits toward an asymptotic regime,  is perturbed by the mass term to be  $g_{\mu\nu} p^\mu p^\nu =-m^2 \equiv -\epsilon$.
 For small perturbations, the scattering orbit will be perturbed  and result in the slightly change in the deflection angle.
Then, in the semiclassical approximation,  the phase shift in (\ref{pws}) can be related to the  deflection angle of the scattering orbits and used to compute the scattering amplitude.
It turns out that the semiclassical results with a small spatially independent mass perturbation qualitatively agree with the full numerical studies in which the mass term is introduced to be space-dependent and vanishes  in an asymptotic regime.
This semiclassical analysis  provides an alternative interpretation of the scattering cross section in terms of the orbiting.

The geodesics follows the spacetime with the line element of the form
\begin{align}
ds^2 =& -\left( 1-\frac{C^2+D^2}{r^2}\right) dt^2 +\left( 1
-\frac{D^2}{r^2}\right)^{-1} dr^2 \nonumber\\
&-2C d\phi dt +r^2 d\phi^2,
\end{align}
where  $C=\ell/m_a c$ for the circulation of the vortex and $D=d/m_a c$ for the draining rate.
The metric is independent of $t$ and $\phi$. There are two killing vectors leading to two conserved quantities, namely energy $E$ and azimuthal angular momentum $L$, along a geodesic. In terms of the proper time $\tau$  \cite{Hsieh2021,Wang2022,Hsieh2025}, they are expressed by
\begin{subequations}
\begin{align}
E &= \left( 1-\frac{C^2+D^2}{r^2}\right) \frac{dt}{d\tau} +C \frac{d\phi}{d\tau}, \\
L &= -C \frac{dt}{d\tau} + r^2 \frac{d\phi}{d\tau},
\end{align}
\end{subequations}
giving two equations of motion
\begin{subequations}
\begin{align}
\frac{dt}{d\tau} &= \frac{E-CL/r^2}{\left( 1-{D^2}/{r^2}\right)} ,\\
\frac{d\phi}{d\tau} &
= \frac{L}{r^2} + \frac{C E-L C^2/r^2}{r^2 \left( 1-{D^2}/{r^2}\right)}.
\end{align}\label{phi}
\end{subequations}

For the geodesics obeying $g_{\mu\nu} p^\mu p^\nu =-m^2 \equiv -\epsilon$ where $p^\mu=m d x^\mu / d\tau$ is the four-momentum, the radial motion can be derived as
\begin{equation}
\begin{split} \label{r}
\left( \frac{dr}{d\tau} \right)^2
=& \left( E-\frac{C L}{r^2} \right)^2 -\left( 1-\frac{D^2}{r^2} \right) \frac{L^2}{r^2} -\epsilon \, \left( 1-\frac{D^2}{r^2}\right) \\
\equiv& \left( \frac{dr^{(0)}}{d\tau} \right)^2 -\epsilon \, \left( 1-\frac{D^2}{r^2}\right) .\\
\end{split}
\end{equation}
Here, we treat the mass term as a perturbation, that perturbs the null geodesics $r^{(0)}$.
Hence, together with (\ref{phi}) and (\ref{r}), the equation  for radial motion up to order $\epsilon$ becomes
\begin{equation}
\frac{dr}{d\phi} = \frac{dr^{(0)}}{d\phi} -\frac{\left( 1-{D^2}/{r^2}\right)}{2 ({dr^{(0)}}/{d\tau})} \frac{1}{{d\phi}/{d\tau}} \epsilon +\mathcal{O}(\epsilon^2) ,
\end{equation}
where the integral form of the deflection angle $\Theta$ shown in Fig.~\ref{Schematic} can be written as
\begin{align} \label{def_angle}
\Theta
&= -\pi +2 \int_{r_t}^\infty \left|\frac{d\phi}{dr}\right| dr \nonumber \\
&= -\pi +2\int_{r_t}^\infty \frac{d\phi}{dr^{(0)}} dr \nonumber
\\
&\quad-2 \epsilon \int_{r_t}^\infty \frac{ \left( 1-D^2/r^2\right) ({d\phi}/{d\tau})}{2 ({dr^{(0)}}/{d\tau})^3}  \, dr +\mathcal{O}(\epsilon^2)\, .
\end{align}
The radial coordinate $r_t$ is a turning point of the trajectory determined by $d r^{(0)}/d\tau=0$ to be
\begin{equation}\label{r_b_sc}
C^2 b^2-2 s C b r_t^2 E +D^2 b^2 -r_t^2 \left(b^2 -r_t^2 E^2 \right) = 0 \,
\end{equation}
{with the impact parameter $b=\frac{\vert L \vert}{E}$ with $s=+/-$ for corotating/counterrotating  orbits.}

The radial velocity above for  null geodesics  takes the form
\begin{equation} \label{velocity_r}
\frac{dr^{(0)}}{d\tau} \sim \frac{1}{r^2} \sqrt{(r-r_1)(r-r_2)(r-r_3)(r-r_4)} \,,
\end{equation}
where we label the roots as $r_4 \ge r_3 >r_2>r_1$, and the detailed expressions of the four roots can be seen in Appendix A in Ref.~\cite{Wang2022}}. For the unstable circular motion,  the radius $r_{sc}$ is given by  the  double root $r_{sc}=r_3 = r_4$ { with the corresponding critical impact parameter defined via \eqref{r_b_sc}.}

It is straightforward  to find the critical impact parameter $b_{sc}$ given by \cite{Hsieh2021,Hsieh2025}
\begin{equation}
b_{\pm c} = 2 \sqrt{D^2 +C^2} \mp 2C,
\end{equation}
and the unstable circular orbits
\begin{equation}
r_{\pm c} = \left(\sqrt{D^2 +C^2} |b_{\pm c}| \right)^{1/2}
\end{equation}
in Ref.~\cite{Dolan2013b}.
Here we primarily focus on the large energy scattering  when the impact parameter  approaches the critical value,  $b \rightarrow  b_{sc}  \propto r_E =\sqrt{D^2 +C^2}$  in the so-called strong deflection limit (SDL) \cite{Hsieh2021,Hsieh2025}.
The leading behavior of the deflection angle in the SDL can be extracted from  the integral in (\ref{def_angle}) in the region of $ r \rightarrow  r_t=r_{sc}$.  Given by the inverse of the radial velocity in (\ref{velocity_r}),  in the case of the double root $r_{sc}=r_4=r_3$,  the first integral behaves like $1/(r-r_{sc})$
and the second integral of the $\epsilon$ correction  results in the form  $1/(r-r_{sc})^3$ , giving the divergence at $r=r_t=r_{sc}$.
 Through the expansion of $b(r_t)$ around $r_t=r_{sc}$,
\begin{equation}
b(r_t) = b_{sc} +\frac{1}{2!} b_{sc}'' (r_t-r_{sc})^2 +\mathcal{O}(r_t-r_{sc})^3 ,
\end{equation}
where $b_{sc}'' \equiv b''(r_{sc})$ and the prime means the derivative with respect to $r_t$,
the divergence behavior can be translated into the form of $\log{(b-b_{sc})}$ and $(b-b_{sc})^{-1}$.
As a result, the deflection angle in the SDL in terms of the impact parameter $b$ is expressed as
\begin{equation}
\begin{split}\label{Theta}
    \Theta^s(b)
    &\approx -\bar{a}_s \log{(b-b_{sc})} +\bar{b}_s +\frac{\epsilon}{E^2} \, \frac{\hat{a}_s}{(b-b_{sc})} \,,\\
\end{split}
\end{equation}
 with  the coefficients
\begin{align}
\label{bara}
    \bar{a}_s =& \frac{r_{sc}}{2 D} \left( \sqrt{1+\frac{C^2}{D^2}} +\frac{s C}{D} \right) ,\\
    \bar{b}_s =& -\frac{r_{sc}}{2 D} \left( \sqrt{1+\frac{C^2}{D^2}} +\frac{s C}{D} \right) \log{\left(\frac{r_{sc}^2}{64 b_{sc} (D^2+C^2)}\right)}\nonumber\\
    & -\frac{s C}{D} \log{\left(\frac{r_{sc}+D}{r_{sc}-D}\right)}  -\pi,
\end{align}
in Ref.~\cite{Dolan2013b}, and
\begin{equation}\label{hata}
\hat{a}_s = \frac{r_{sc}^3 \left[-C^2 b_{sc} +s C r_{sc}^2 +b_{sc} \left(r_{sc}^2 -D^2 \right)\right]}{2  (6 C^2 b_{sc}^2 -2 s C b_{sc} r_{sc}^2 +6 D^2 b_{sc}^2 -b_{sc}^2 r_{sc}^2)^{3/2}}\, .
\end{equation}

In Fig. \ref{Schematic}, the deflection angle for the corotating orbit is given by
\begin{equation}\label{Theta+}
    \Theta^+(b) = 2n\pi +\phi ,
\end{equation}
and for the counterrotating  is in particular defined as
\begin{equation}\label{Theta-}
    \Theta^- (b)= 2n\pi -\phi ,
\end{equation}
where $0<\phi<2\pi$.
Note that $n=0$ ($n=1$) is for the corotating (counterrotating ) orbit winding around the black hole zero time, which are two  orbits contributing significantly to the semiclassical differential cross section to be discussed later \cite{Dolan2013b}. One can write $ b(\phi)$ by the inversion of (\ref{Theta}) and with the deflection angle given by (\ref{Theta+}) and (\ref{Theta-}) as
\begin{equation}
    b(\phi) = b_{sc} +b^{(0)}_s(\phi) + \frac{\epsilon}{E^2} \, b^{(1)}_s(\phi),
\end{equation}
where
\begin{equation}
    b^{(0)}_\pm(\phi) = \exp\left[ \frac{-\bar{b}_\pm+2n \pi \pm \phi }{-\bar{a}_\pm} \right],
\end{equation}
and its correction of  $\epsilon$ is{
\begin{equation}
    b^{(1)}_\pm(\phi) \approx \frac{ \hat{a}_\pm}{\bar{a}_\pm}.
\end{equation}
}

The classical differential  cross section of the scattering is determined by the deflection angle
\begin{equation}
    \frac{d\sigma}{d\phi} \bigg|_{\text{cl}} = \bigg| \frac{d\Theta}{db} \bigg|^{-1}.
\end{equation}
Quantum mechanically, the differential cross section can be computed  from the scattering amplitude as in \eqref{d_cross_sec},
where the scattering amplitude is given by  the sum of the contributions of the scattered wave functions through their the phase shifts
\begin{align}\label{f}
	f_\omega(\phi)
                  &\approx \sqrt{\frac{1}{2i\pi\omega}}\sum_{m=-\infty}^{\infty}  e^{2i\delta_m} e^{im\phi}
\end{align}
 in which the on-axis contributions are ignored \cite{Dolan2013b}.
 The expression can be converted to the integral by the Poisson sum formula \cite{Nussenzveig1992} as
 \begin{align}
 f_\omega(\phi)=&\sqrt{\frac{1}{2i\pi\omega}}\sum_{n=-\infty}^{\infty} e^{-i\pi n} \int_{-\infty}^{\infty} dm \, e^{2i\delta_m} e^{im\phi} e^{2i\pi mn}\nonumber\\
  =&\sqrt{\frac{1}{2i\pi\omega}}   \sum_{n=-\infty}^{\infty} \bigg[ e^{-i\pi n} \int_{-\infty}^{\infty} dm \, e^{2i\delta_m} e^{im\phi} e^{2i\pi mn}
 \nonumber\\&+ e^{i\pi n} \int_{-\infty}^{\infty}dm \,  e^{2i\delta_{-m}} e^{-im\phi} e^{2i\pi mn}\bigg].
 \end{align}
 Let us define
\begin{equation}
\xi^\pm_n = 2 \delta_{\pm m(\phi)} \pm m \phi +2 \pi n m(\phi)
\end{equation}
in terms of the phase shift $\delta_{\pm m}$ and $m(\phi)$, which can be linked to the orbital dynamics as $m(\phi) =\omega b(\phi)$ for fixed $n$.
In the stationary phase condition,
\begin{equation}
\frac{d}{dm}\xi_n^\pm  =\frac{d}{dm}[2\delta_{\pm m}] \pm \phi+2n\pi =0 \, .
\end{equation}
Then the phase shift can be related to  the deflection angle in (\ref{Theta+}) and (\ref{Theta-}) as
\begin{equation}\label{delta_Theta}
\Theta^\pm = - \frac{d}{dm} (2\delta_{\pm m})\, ,
\end{equation}
and it can be analytically computed by integrating the deflection angle over $m=\omega b$. With the expression of the deflection angle in (\ref{Theta}), the phase shift in the leading order of the SDL  becomes
{
\begin{equation}
\begin{split}
\delta_{\pm m} &= -\frac{\omega}{2} \int \Theta^\pm(b) \, db \\
&\approx -\frac{\omega}{2} \bigg\{ -\bar{a}_\pm (b-b_{\pm c})\log{(b-b_{\pm c})} +\bar{a}_\pm b +\bar{b}_\pm b \nonumber\\
&+\frac{\epsilon}{E^2} \, \hat{a}_\pm \log{(b-b_{\pm c})} \bigg\}.
\end{split}
\end{equation}}
Let us consider the contributions to the differential scattering cross section  from two dominant contributions, $n=0$ for the corotating orbit and $n=1$ for the counterrotating  orbit.
The total differential cross section can be approximated by the sum of the respective classical differential cross sections  and their interference effect given by
\begin{equation}
    \frac{d\sigma}{d\phi} \approx \frac{d\sigma^+}{d\phi} + \frac{d\sigma^-}{d\phi} + I,
\end{equation}
where{
\begin{equation}\label{dtheta}
    \frac{d\sigma^\pm}{d\phi} = \bigg| \frac{d\Theta^\pm}{db} \bigg|^{-1}
    \approx \bigg| \frac{-\bar{a}_\pm}{b-b_{\pm c}} +\frac{\epsilon}{E^2} \, \frac{- \hat{a}_\pm}{(b-b_{\pm c})^{2}} \bigg|^{-1},
\end{equation}
}
and the interference term $I$ between the two orbits is thus
\begin{equation}
{I = 2 \bigg| \frac{d\Theta^+}{db} \frac{d\Theta^-}{db} \bigg|^{-1/2} \cos{(\xi^+_0-\xi^-_1)}}\, .\label{I}
\end{equation}
\begin{figure}[t]
	\centering
	\includegraphics[width=1\linewidth]{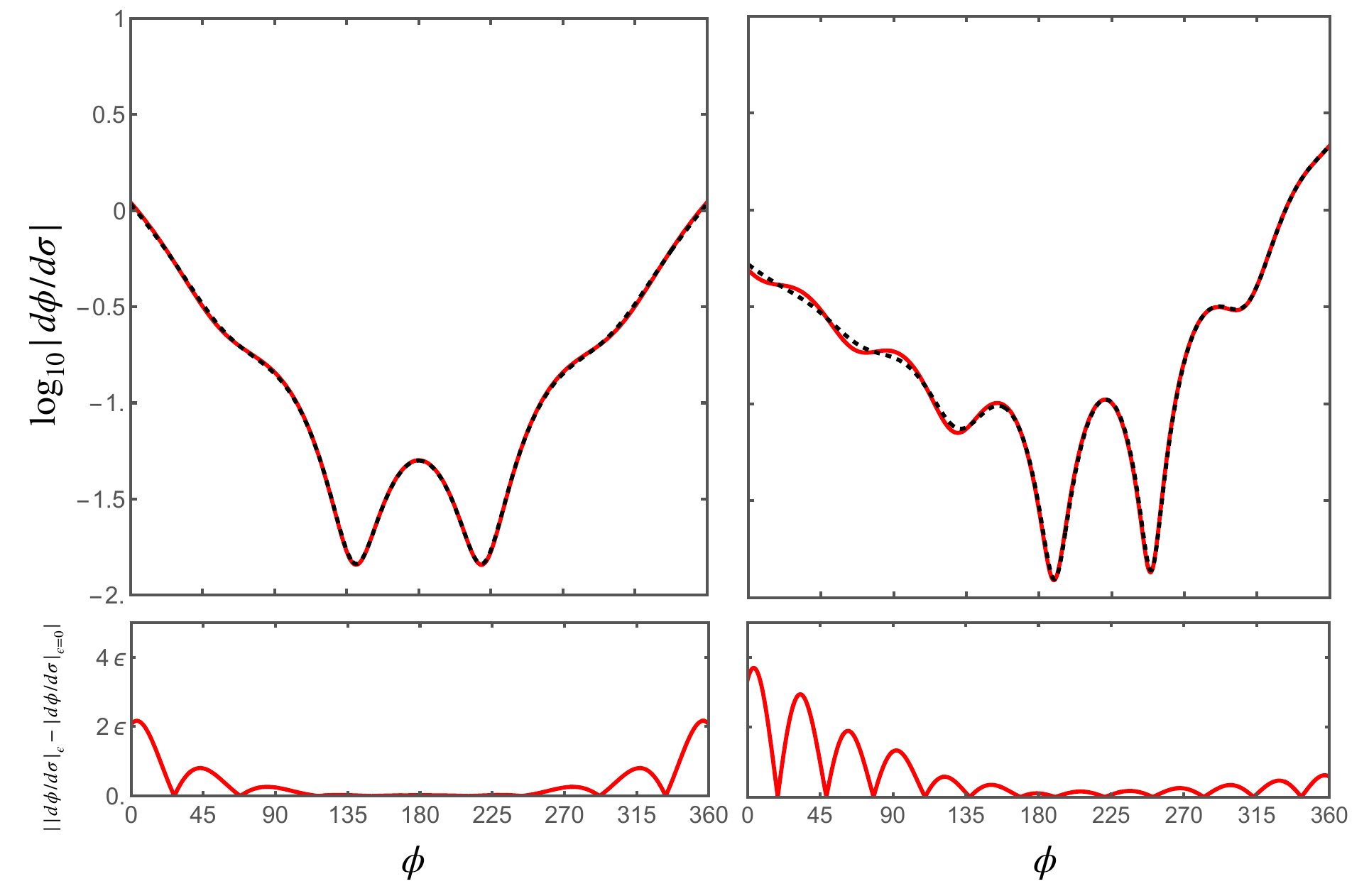}
	\caption{Differential cross sections $d\sigma/d\phi$ as a function of scattering angle $\phi$, showing contributions from the corotating orbit ($n=0$) and the counterrotating  orbit ($n=1$). Results are shown for $C=0$ (left) and $C=1$ (right), with frequency $\omega=1$. The deviation of the scattering amplitude due to the
		perturbations as a function of the scattering angle is plotted. }\label{diff_cross_semiclass}
\end{figure}
 Fig. \ref{diff_cross_semiclass} shows the deflection cross section in (\ref{I}) with/without perturbations for nonrotating/rotating backgrounds.
 Based on (\ref{dtheta}), for the
corotating  (counterrotating ) orbit, the deflection angle of $0 < \phi  < \pi$ ($\pi < \phi  <2\pi$) is  with the impact parameter $b \gg b_{+c}$ ( $b \gg b_{-c}$) where the inverse of $\frac{d\Theta^+}{db}$  ($\frac{d\Theta^-}{db}$ ) is
large that dominates the total cross section at $0 < \phi  <\pi$ ( $\pi < \phi  < 2\pi$). The interference term induces the oscillations with the wavelength  in terms of $\omega$ and $r_E$  as \cite{Dolan2013b}
\begin{equation}
\begin{split}\label{lambda}
\lambda \approx \frac{\pi}{2\omega r_E} - \frac{\epsilon}{E^2} \, \left(\frac{\hat{a}_-}{\bar{a}_-}+\frac{\hat{a}_+}{\bar{a}_+} \right) \frac{\pi}{8 \omega^2 r_E^2}\, ,
\end{split}
\end{equation}
where the mass perturbation shifts the wavelength of order $\epsilon$ also seen in Fig.~\ref{diff_cross_semiclass}. The detailed derivation can be seen in  Appendix B.
Apart from that, the mass perturbation results in  the relatively large modification of the deferential cross section  at the relatively small scattering angle due to the interference term $I$.  This can be understood by the fact that the interference term involves the inverse of $\frac{d\Theta^+}{db}$
for the co-rotation orbit and $\frac{d\Theta^-}{db}$
for the counter-rotation orbit. At the scattering angle of $0 < \phi < \pi$ ( $\pi < \phi < 2\pi$), the correction of order mass perturbation $\epsilon$  becomes evident arising from  the large value of the inverse of the unperturbed $\frac{d\Theta^+}{db}$ ($\frac{d\Theta^-}{db}$) as $ b \gg b_{+c}$ ( $ b \gg b_{-c}$). In addition,  in the rotating background with $C=1$, the effect of the perturbations to the scattering amplitude is more profound at the  angle, of $0 < \phi < \pi$ as compared with the angle of $\pi < \phi < 2\pi$. In the expression of  (\ref{dtheta}), the perturbed effect of $\frac{d\Theta}{db}$ is controlled by $\hat a/\bar{a}$ with $\bar a$ and $\hat a$ given in (\ref{bara}) and (\ref{hata}) respectively. In general, $\bar a_- < \bar a_+$  due to $b_{-c} >b_{+c}$, which has been studied in   Ref.~\cite{Hsieh2021} and it is also found that $\hat a_- > \hat a_+$, so that they lead to the large perturbed effect to the perturbed part of $\frac{d\Theta^-}{db}$. Through the interference term $I$, the perturbed effect is amplified again due to the largeness of the inverse of the unperturbed  $\frac{d\Theta^-}{db}$  at the angle of $0 < \phi < \pi$.

The cross section in the stationary  phase approximation provides the qualitative feature of its full numerical result. In the numerical studies, the mass perturbation will be introduced to be space-dependent and then trigger the destabilization of the scattering amplitudes.

\section{Regge Poles}\label{seciv}
In addition to the  semiclassical treatment with a constant perturbation term, introducing  a spatially dependent bump in the effective potential  given by \eqref{Veff2} allows us to model a more realistic situation in which a black hole is surrounded by environment matter.
 We numerically investigate the wave equation \eqref{WEQ}, where the solution  obeys the boundary conditions expressed in (\ref{ingoing}).
By further assuming that only the outgoing wave exists at infinity leads to  $A^\text{in}=0$ in \eqref{ingoing} where one can have
 a spectrum of discreet complex-valued frequencies  with a real-valued azimuthal numbers, the QMN, or complex-valued  azimuthal numbers $m$ with the real-valued frequencies, the RPs.

In scattering theory, the RPs can be used together with  the CAM method to compute the scattering amplitude, as an alternative to the partial wave summation approach \eqref{pws}. Through the Watson transformation \cite{Newton1982}, the partial wave summation in (\ref{f}) can be reformulated as a contour integral in the complex-$m$ plane \cite{Dolan2013b}
\begin{align}
	f_\omega(\phi)=&\sqrt{\frac{1}{2i\pi\omega}}\bigg[\int_{-\infty}^{\infty}S(m)e^{im(\phi-2\pi)}dm \nonumber\\
	 &+\frac{i}{2}\int_{-\infty+i\zeta}^{\infty+i\zeta}\frac{S(m)e^{im(\phi-\pi)}}{\sin{m\pi}}dm\nonumber\\
	 &-\frac{i}{2}\int_{-\infty-i\zeta}^{\infty-i\zeta}\frac{S(m)e^{im(\phi-3\pi)}}{\sin{m\pi}}dm\bigg],
	\label{contour_integral}
\end{align}
where $\zeta$ is a small real number, and $S(m)=ie^{im\pi}A^\text{out}/A^\text{in}$, with the poles when $A^\text{in}=0$. For high frequencies, one can neglect the  background contribution in the first term due to the absence of stationary phase point \cite{Dolan2013b}  and consider the contour enclosed in the upper half-plane for the second term, and in the lower half-plane for the last term, respectively. Then 	 \eqref{contour_integral} can be further expressed as the sum of the residues (${\rm Res}$) of the RPs as
\begin{align}
	f_\omega(\phi)\approx&-\sqrt{\frac{\pi}{2i\omega}}\sum_{n=0}^\infty\bigg[\frac{e^{im_{+n}(\phi-\pi)}}{\sin{\pi m_{+n}}}\Res_{m\rightarrow m_{+n}} S(m)\nonumber\\
	&+\frac{e^{im_{-n}(\phi-3\pi)}}{\sin{\pi m_{-n}}}\Res_{m\rightarrow m_{-n}} S(m)\bigg].
\end{align}
Hence we can investigate the individual RP contributions to the scattering amplitude. It will be found below that summing up the contributions of sufficiently many dominant RPs can approximately reach the scattering amplitude by the partial wave summation.

\subsection{Regge pole spectrum and  stability criterion against perturbations}

\begin{figure}[t]
	\includegraphics[width=\linewidth]{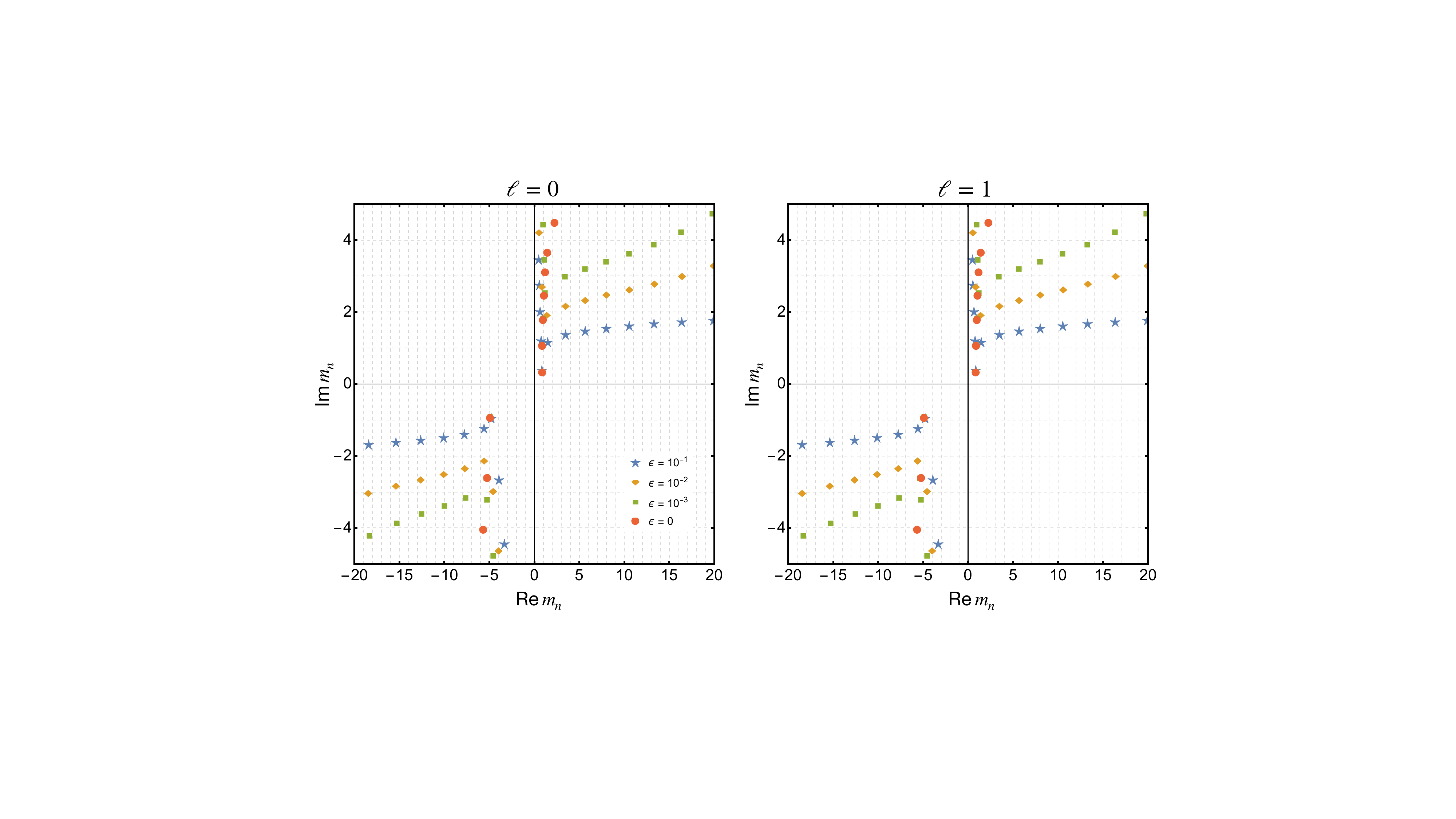}
	\caption{The RP spectra are  shown under various perturbations in the nonrotating background ($\ell=0$) in the left panel, and the rotating background ($\ell=1$) in the right panel. The parameters are fixed at a frequency $\omega=1.0$, bump position $r_0=30$, and width $\alpha=3/2$.}
	\label{fig_ReggePoles}
\end{figure}
We first compute the RPs with or without inhomogeneous mass perturbation (\ref{mass}) in both the nonrotating ($\ell=0$) and rotating ($\ell=1$) backgrounds using a modified continuous fraction method (see Appendix \ref{appA}).
The resulting RP spectrum is shown in a complex-$m$ plane in Fig.~\ref{fig_ReggePoles}, where the filled red circles denote the unperturbed RPs, while other symbols correspond to the RPs  with various values of the perturbations $\epsilon$.

For $\ell=0$, the spectrum exhibits symmetry between the corotating modes located at the upper right quadrant and counterrotating  modes  at the lower left quadrant,namely $m_{\pm,n}^{(\epsilon)}=-m_{\mp,n}^{(\epsilon)}$, while for $\ell \neq 0$, the background rotation  breaks this symmetric feature. The fundamental mode ($n=0$) corresponds to the smallest absolute value of the imaginary part of $m$, i.e., $\vert \text{Im}\,m\vert$.
Upon turning-on perturbations, the RP spectrum begins to bifurcate into two distinct branches: a inner branch with the poles characterized by smaller values of $\vert \text{Re}\, m_\omega\vert$  and  an outer branch with the poles of  the larger values of $\vert \text{Re}\, m_\omega\vert$.
According to \cite{Torres2023}, the bifurcation in the RP spectrum is regarded as a sign of RP spectrum destabilization caused by the perturbations.

Figure~\ref{fig_ReggePoles} shows that in general, the onset of bifurcation is located at the higher overtone given by  smaller $\epsilon$ perturbations and moves toward the fundamental RP with relatively larger $\epsilon$ perturbations for both co- and counter- rotating modes. In the case of $\ell=1$ of the rotating background, the bifurcation with $\epsilon=10^{-3}$ starts  between the third (second) overtone and  the forth (first) overtone for the corotating (counterrotating ) modes and as the perturbation $\epsilon$ increases to  $\epsilon=10^{-1}$, it
 shifts toward the first overtone (fundamental RP).
Thus, the  corotating modes $m_{+,n}$ are more resistant to perturbations and tend to remain stable whereas the counterrotating  modes $m_{-,n}$  become susceptible to perturbations, to the extent that the onset of bifurcation shifts toward the fundamental RP, resulting in destabilization.
As the perturbation $\epsilon$ increases,  the fundamental tone can become destabilized at a certain critical $\epsilon$ to be studied later.
As in \cite{Lucas2025}, which studied the QNMs in the associated pseudospectrum, it was also found that the prograde overtones become more stable than the retrograde overtones when they are perturbed  in the rotating background.

For a fixed frequency $\omega$,  we quantitatively evaluate the modification of the RPs under the perturbations given by
\begin{align}
	\bigg\vert\frac{ \lambda_{\pm,n}^{(\epsilon)}-\lambda_{\pm,n}^{(0)}}{\lambda_{\pm,n}^{(0)}} \bigg\vert_\omega=\bigg\vert\frac{\Delta \lambda_{\pm,n}^{(\epsilon)}}{\lambda_{\pm,n}^{(0)}} \bigg\vert_\omega,\qquad n=0,1,2,3\cdots
	\label{ratio}
\end{align}
with $\lambda_{\pm,n}^{(\epsilon)}=m_{\pm,n}^{(\epsilon)}l$.
Following  the criterion in  \cite{Cheung2022},   the RP is considered  as destabilized when the ratio $\vert\Delta\lambda_{\pm,n}^{(\epsilon)}/\lambda_{\pm,n}^{(0)}\vert$ is much larger than the perturbation $\epsilon$. Conversely, when the ratio remains smaller than or the same order of $\epsilon$, the poles are  regarded as stable.

\subsection{Migration of  RP of overtones}
\begin{figure*}[th]
	\includegraphics[width=\linewidth]{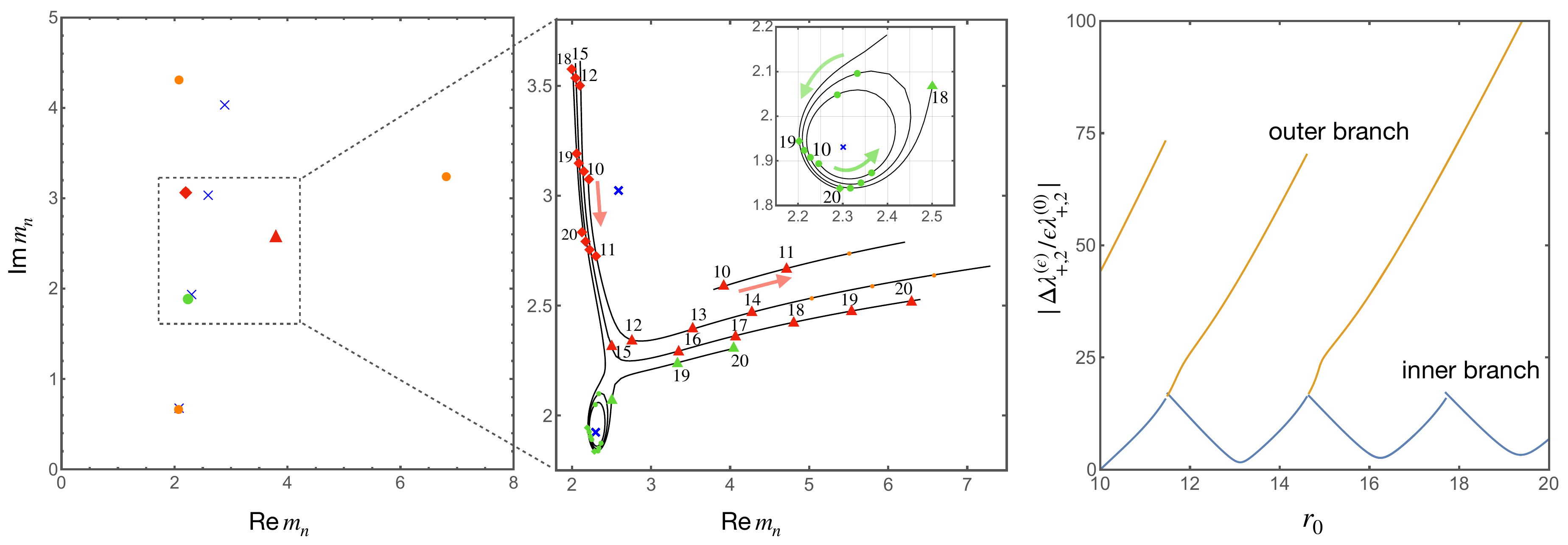}
	\caption{  Migration behavior for RPs  illustrated by the case of $\ell=0$ for high frequency $\omega=1.0$ as the bump position increases  from $r_0=10$.  In the left panel, we highlight the RP of the first overtone (blue cross) and the RPs  of the bifurcated second overtone in the outer branch (red triangle) and inner branch (red square). In the middle panel, as $r_0$ increases, RPs have large migration or overtaking jump (as shown in the inset) signaling the shift of the bifurcation from the second overtone to the first overtone with the large $\bigg\vert\frac{\Delta \lambda^{(\epsilon)}}{\lambda^{(0)}} \bigg\vert_\omega$  in the right panel. Note that the spacing between adjacent red dots in the right panel is $\Delta r_0=1$.}
	\label{fig_traj}
\end{figure*}
 To further explore the bifurcation spectrum, we consider the migration of the PRs by changing  the bump position $r_0$ in \eqref{mass} and focus on the large migration and the discontinuous overtaking jumps, both of which serves as  strong indicators of destabilization \cite{Torres2023, Cheung2022}.

 We zoom in on the poles of the second overtone of the corotating modes including   the poles  on the inner and outer branches as well as the first overtone in the left panel of Fig.~\ref{fig_traj} and track the trajectories of the poles starting from $r_0=10$.
 As the bump position increases from $r_0=10$ to $20$, the pole of the second overtone in the outer branch reveals several overtaking jumps at approximately $r_0=12, 15$. The poles in the inner branch also have overtaking jumps at approximately $r_0=12, 15$, and $18$.
  As for the first overtone, the spiral trajectory of the pole of the first overtone starts to bifurcate into two branches at $r_0=18$.

Remarkably, the  onset of bifurcation shifts  to the first overtone.
Note that the same migration scheme where the onset of the bifurcation is shifted from the first overtone to the fundamental RP also applies to the counterrotating  modes.
Therefore, the bifurcations and the consequent overtaking migrations in the complex-$m$ plane can be regarded as a signature of RP destabilization with large values of $\bigg\vert\frac{\Delta \lambda_{\pm,n}^{(\epsilon)}}{\lambda_{\pm,n}^{(0)}} \bigg\vert_\omega$, which can be as large as $\bigg\vert\frac{\Delta \lambda_{+,2}^{(\epsilon)}}{\lambda_{+,2}^{(0)}} \bigg\vert_\omega \approx 70\epsilon$ in the most right panel plot of Fig.~\ref{fig_traj} for the outer branch pole. The resulting impact on the scattering amplitude, particularly at large frequency, will be discussed later.

\begin{figure}[b]
	\includegraphics[width=1\linewidth]{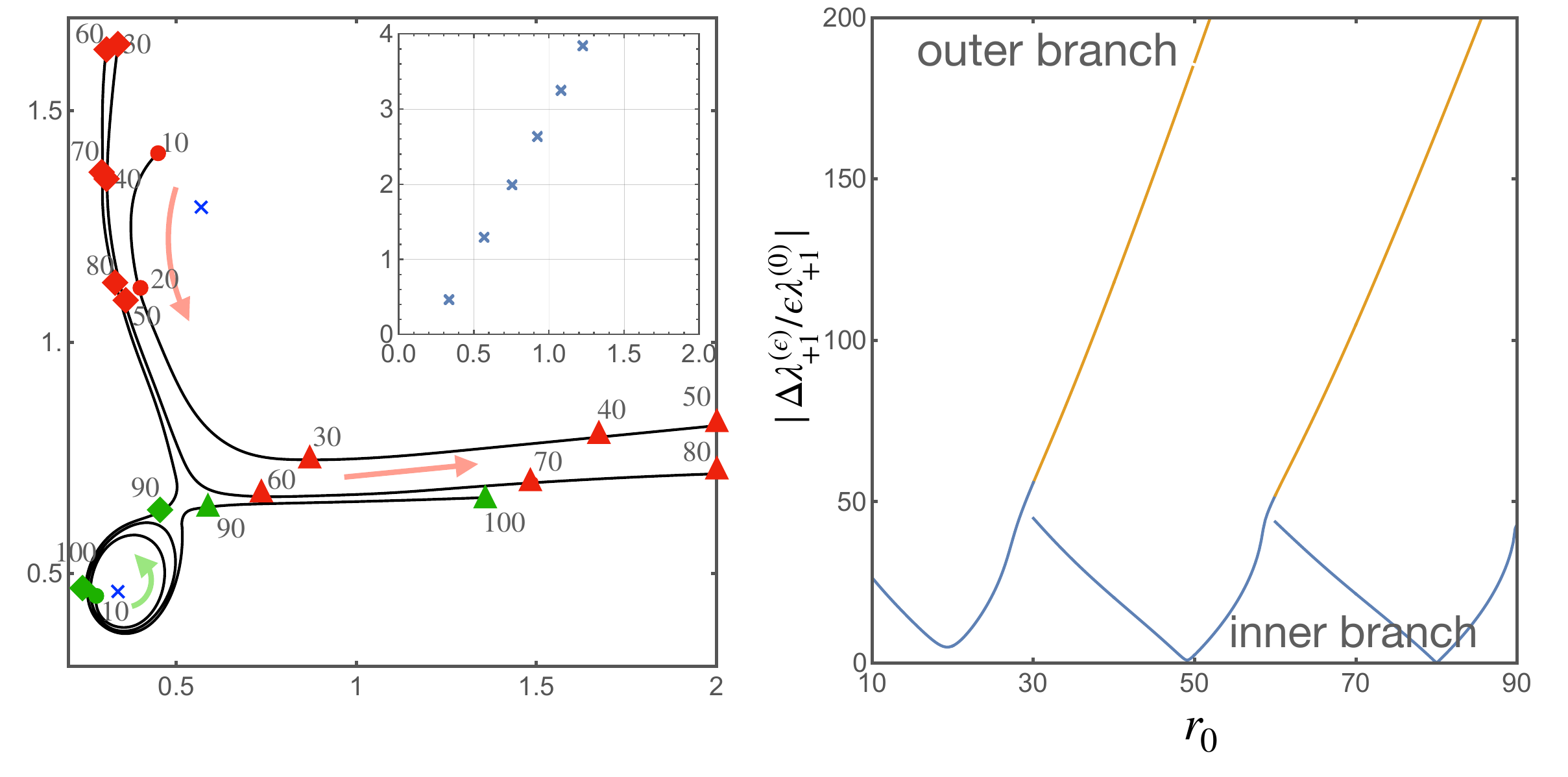}
	\caption{   Migration behavior for RPs  illustrated by the case of $\ell=0$ for low frequency $\omega=0.1$ as the bump position increases from $r_0=10$.  Migration of the RP of the fundamental poles (blue triangle/square)  and the RPs  of the bifurcated first overtone in the outer branch (triangle red) and inner branch (square red), signaling the shift of the bifurcation from the first overtone to the fundamental pole. The inset shows the unperturbed RPs spectrum. The right panel is  the values $\vert\Delta \lambda_{+1}^{(\epsilon)}/\epsilon\lambda_{+1}^{(0)}\vert$ for both the inner and outer branches as a functions of $r_0$.} 
	\label{fig_traj_lowf}
\end{figure}

As for the low frequency case shown in Fig.~\ref{fig_traj_lowf}, the bifurcation happens in the first overtone at $r_0=30$. Then, when $r_0$ increases to $r_0=90$, the pole in the outer branch undergoes overtaking jump again at $r_0=60$ where $\bigg\vert\frac{\Delta \lambda_{+,1}^{(\epsilon)}}{\lambda_{+,1}^{(0)}} \bigg\vert_\omega \approx 50\epsilon$. The pole in the inner branch also has another overtaking jump at $r_0=60$.
At $r_0=90$, the fundamental RP bifurcates with the pole in its outer branch merging with the outer branch pole of the first overtone. From then, the onset of the bifurcation shifts from the first overtone to the fundamental RP. The destabilization of the poles will influence the scattering amplitude at relatively {large} scattering angles to be seen later.

\subsection{Stability of the fundamental RP}
\begin{figure}[t]
	\includegraphics[width=\linewidth]{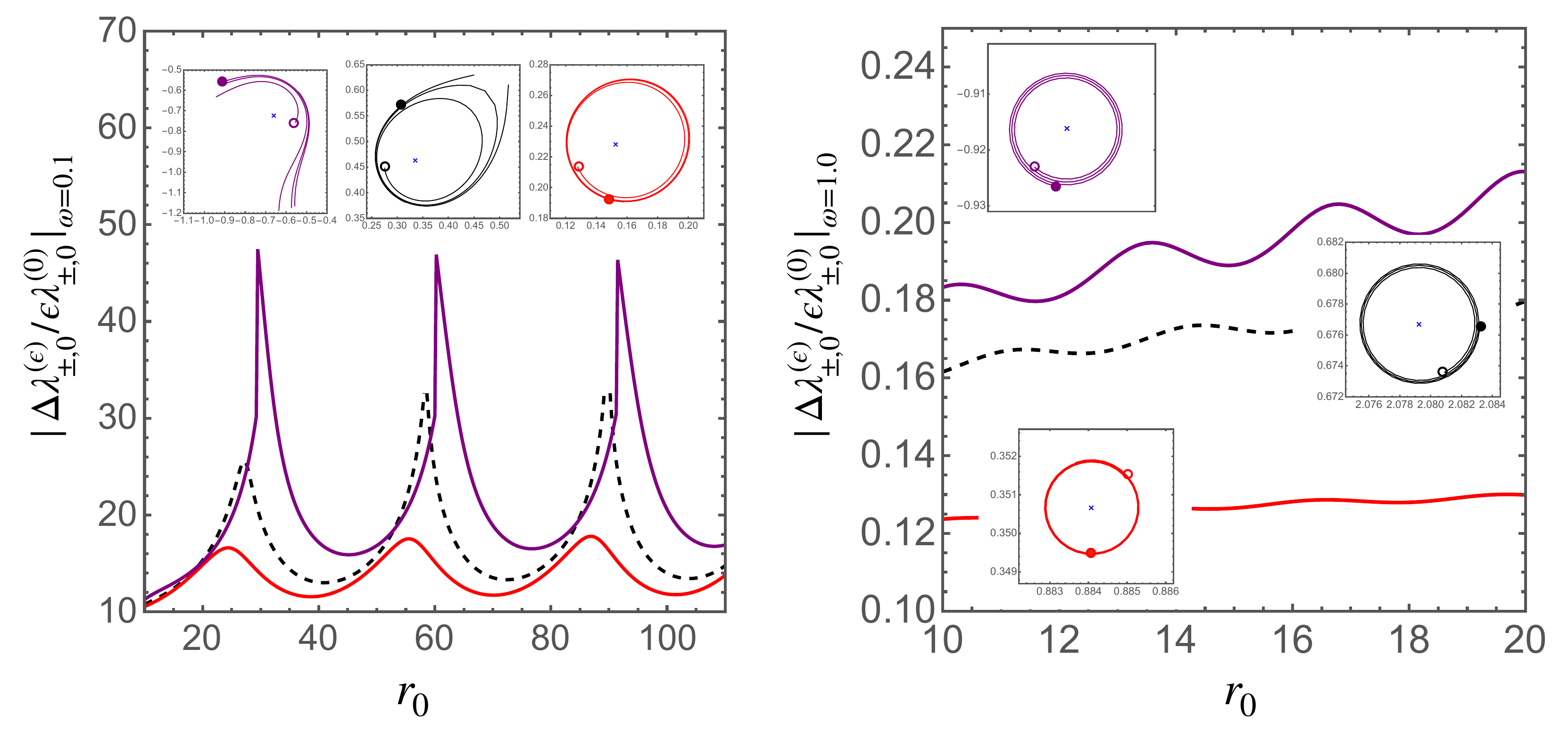}
	\caption{ The plots show the relative deviation
		$\vert \Delta \lambda_{\pm,0}^{(\epsilon)}/\lambda_{\pm,0}^{(0)}\vert$ of the fundamental RP as a function of $r_0$ where the purple (red) line is for the counterrotating   (corotating) case, and the  black dashed line is for the nonrotating background with the
		results  shown for  frequencies $\omega=0.1$ (left panel) and $\omega=1.0$ (right panel). }
	\label{fig_fund}
\end{figure}
We now  focus on the fundamental RPs and track their trajectories in the complex-$m$ plane as the bump position $r_0$ increases from $10$ to $100$, as depicted in Fig.~\ref{fig_fund}.
The figure shows in both nonrotating and rotating backgrounds, at low frequency, the quantity $\vert\Delta\lambda_{\pm,0}^{(\epsilon)}/\lambda_{\pm,0}^{(0)}\vert$ becomes large, indicating either significant migration or overtaking jumps in the pole trajectories. In contrast, at high frequencies, this ratio remains small and stable, and the pole migration follows a closed trajectory.

In particular, in the nonrotating background and at low frequency, the trajectory of the perturbed fundamental RP shows large migration at $r_0\simeq25$ and $55$, followed by an overtaking jump at $r_0\simeq90$, resulting in corresponding peaks of $\vert\Delta\lambda_{0}^{(\epsilon)}/\lambda_{0}^{(0)}\vert$.
In the rotating background and also for low frequency,  the  trajectory of $m_{-,0}^{(\epsilon)}$ for the counterrotating  mode  starts  with the continuously spiral shift, and change to the discontinuously overtaking  jump  at $r_0\simeq30,\,60,$ and $95$. These transitions lead to peaks into $\vert\Delta\lambda_{-,0}^{(\epsilon)}/\lambda_{-,0}^{(0)}\vert$, with values larger than those observed in the nonrotating background.

 Conversely, regardless of the increase in $r_0$, the migration  of $m_{+,0}^{(\epsilon)}$ of the corotating mode  remains in a closed trajectory  with a relatively smaller value of $\vert\Delta\lambda_{+,0}^{(\epsilon)}/\lambda_{+,0}^{(0)}\vert$ than that of the nonrotating case.
Again, for  the fundamental RP in the rotating background, the corotating mode is more stable against the perturbations than the counterrotating  mode.
The destabilization of the fundamental RP will influence the scattering amplitude in particular for low frequency at large scattering angles.
\begin{figure}[t]
	\includegraphics[width=\linewidth]{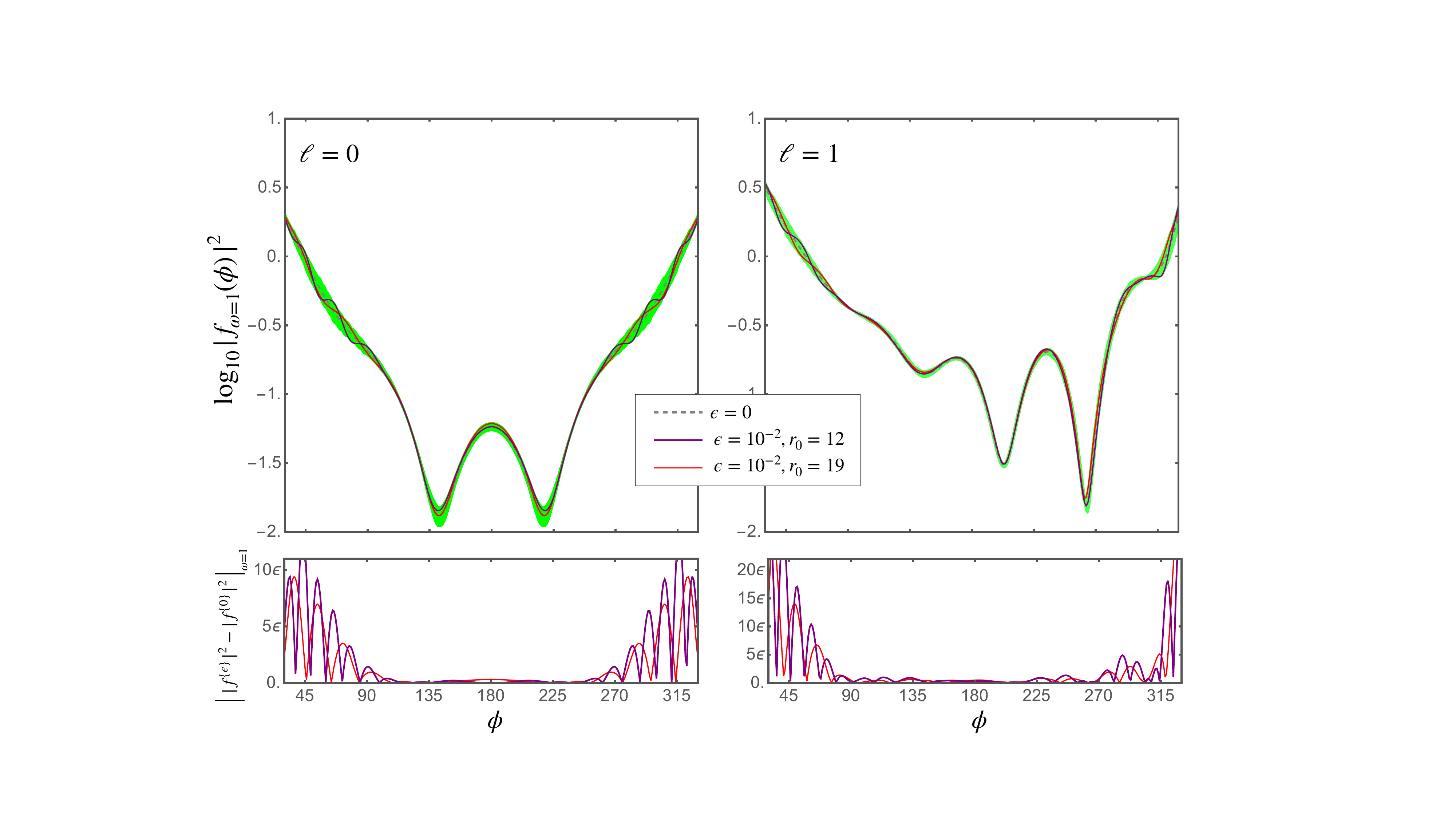}
	\caption{The logarithm of the squared scattering amplitude is shown for  high frequency $\omega = 1.0$ in  the nonrotating  (left) and the rotating (right) backgrounds, including both the unperturbed (blue curve) and perturbed cases.
		As the bump ($\alpha=3/2$, $\epsilon=10^{-2}$) position increases from $r_0=10$ to $20$,  the significant change is shown  by the green shadow at relatively small scattering angles. The deviation of the scattering amplitude due to the perturbations as a function of the scattering angle is plotted.}
	\label{fig_obs_highf}
\end{figure}
\begin{figure}[b]
	\includegraphics[width=0.9\linewidth]{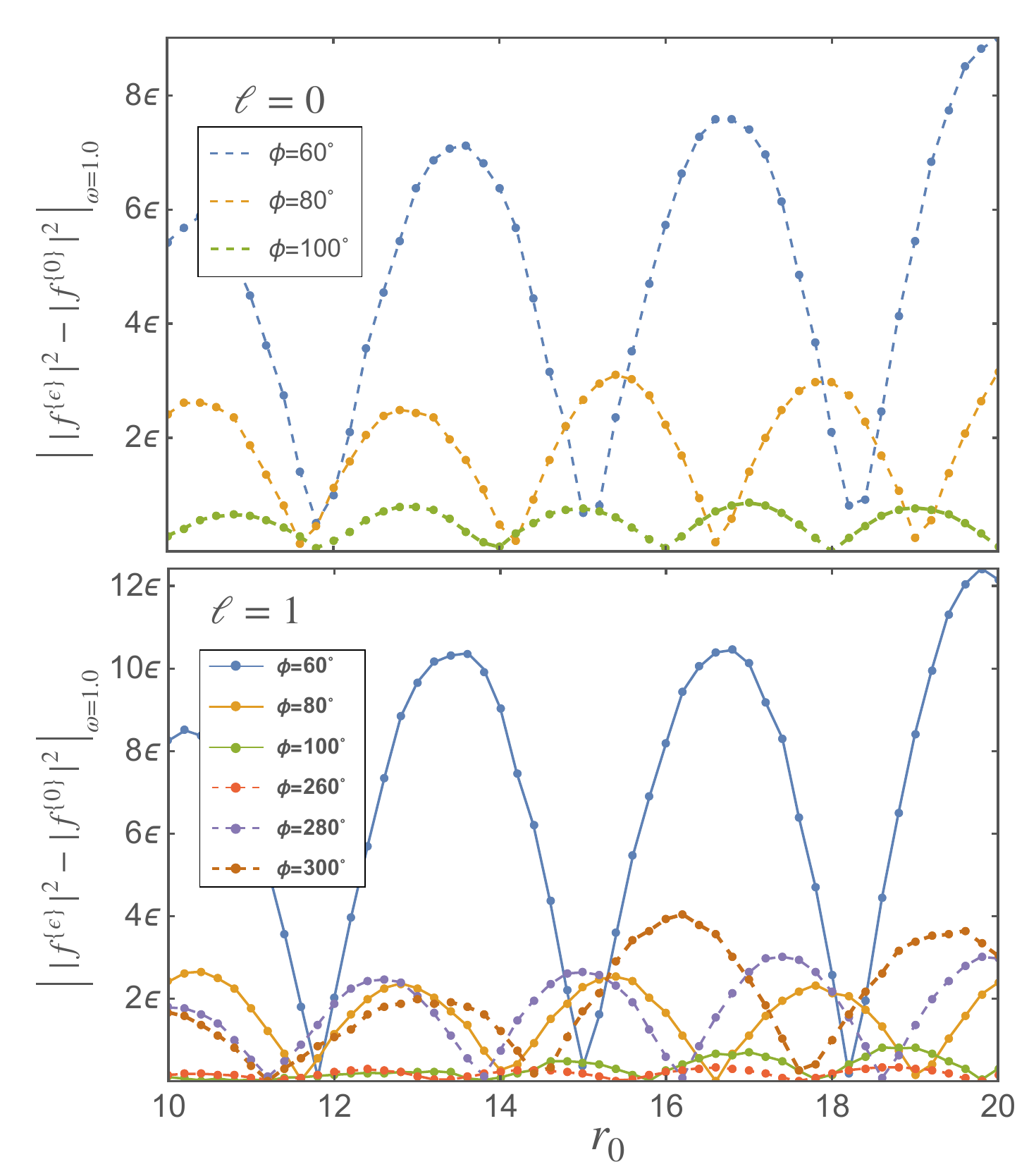}
	\caption{The deviation of the scattering amplitude due to the perturbations for high frequency $\omega=1.0$ in both the nonrotating  (top) and rotating (bottom) backgrounds at various scattering angles. Other parameters: $\alpha=3/2$, $\epsilon=10^{-2}$.}
	\label{fig_obs_angle_highf}
\end{figure}
\begin{figure*}[t]
	\includegraphics[width=0.9\linewidth]{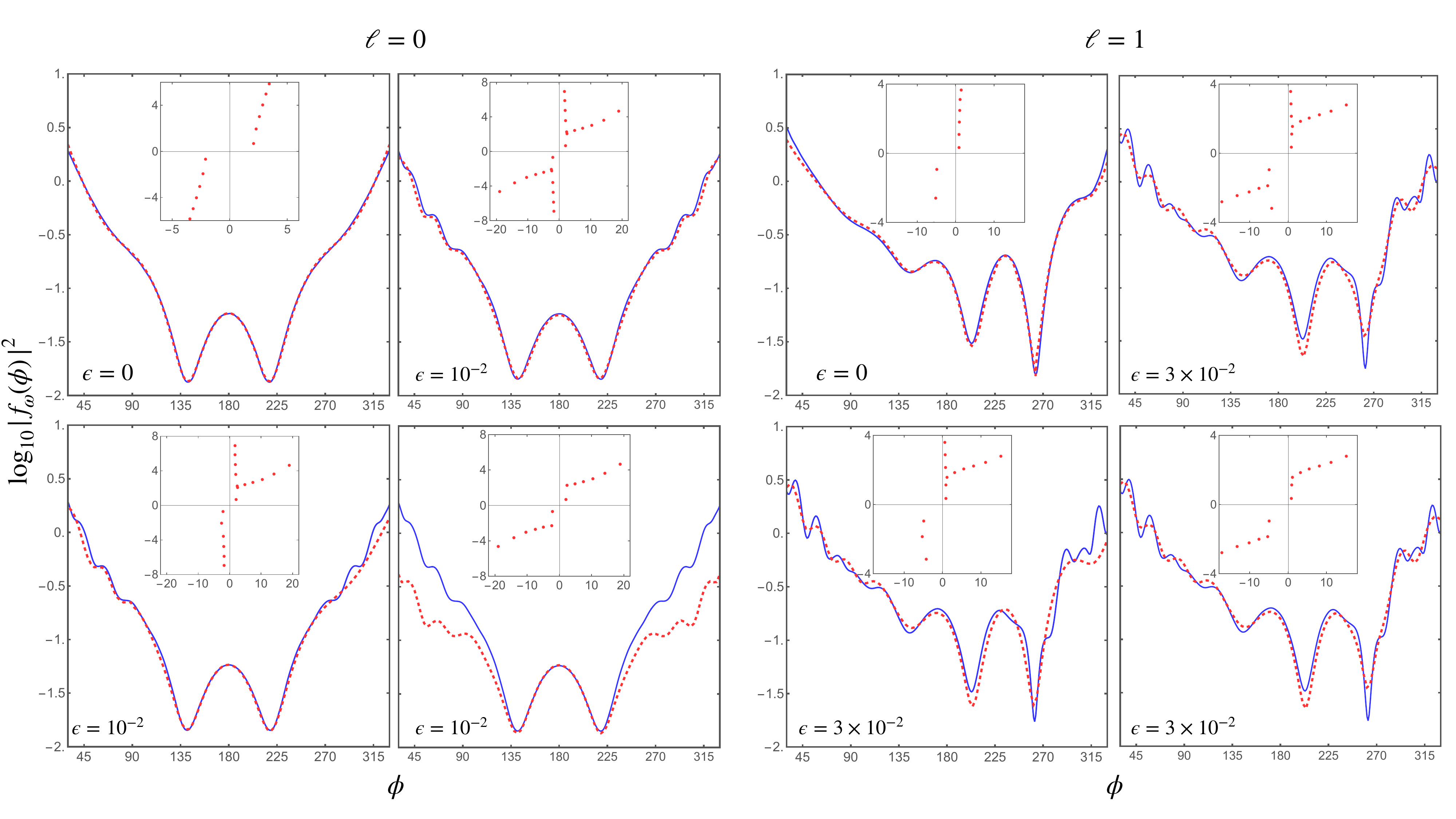}
	\caption{ Distinct pole branches contribute the scattering amplitude in their respective angular regions in both nonrotating and rotating backgrounds  with frequency $\omega=1.0$, bump position $r_0=20$, and width $\alpha=3/2$. In the cases of $\ell=0$ and $\ell=1$, the respective upper left plot  is for the unperturbed scattering amplitude  with the contributions of the RPs shown in the inset whereas the  upper-right one is for  the perturbed scattering amplitude with the contributions of the inner and outer pole branches. The lower-left plot is for the perturbed scattering amplitude with all contributions of the  PRs  but without the outer pole branch for $\text{Re}\, m_\omega <0$. The lower-right plot is for the perturbed scattering amplitude with all contributions of the  PRs  but without the inner pole branch for $\text{Re}\, m_\omega <0$
		and $\text{Re}\, m_\omega >0$.}
	\label{fig_CAM}
\end{figure*}
\section{Numerical study of scattering cross section}\label{secv}
While the destabilized QNM spectrum has already been studied in the  DBT under the environmental effect in Ref.~\cite{Syu2024}, the instability there was demonstrated through  time-dependent observables.
Now  we would like to study the observational consequence of RP destabilization  mainly resulting from the spectrum bifurcation and pole migration by shifting the bump location $r_0$.

To do so, we first numerically employs the partial wave summation method to obtain the scattering amplitude  and the differential cross section  under the perturbations, and compare the results to those of the unperturbed ones.
As presented in Fig.~\ref{fig_obs_highf} for relatively high frequency, the effect of perturbations leads to an amplification of  the oscillation amplitude in the differential cross section at small scattering angles,  which  can be qualitatively understood through semiclassical analysis  due to the interference between the corotating and counterrotating  orbits, as shown in Fig.~\ref{diff_cross_semiclass}.

However,  for inhomogeneous mass perturbations, the amplification  is more effective and is seen resulting from the destabilization of the RPs, where the poles of the overtones in the outer branch reveal the large migration or overtaking jumps at some particular values of $r_0$, as shown in Fig.~\ref{fig_traj}, with which $r_0$,  the scattering amplitude  has  the peak values in Fig.~\ref{fig_obs_angle_highf}.
For $\ell=1$ of the rotating background, the amplification mechanism is even more effective than in the $\ell=0$ case of the nonrotating background  at the scattering angle of $0<\phi <\pi$, which also been seen in the  semiclassical analysis in Fig.~\ref{diff_cross_semiclass}.
Then, we employ the CAM method to try to understand  the contributions of each pole branch to the scattering amplitude.
In Fig.~\ref{fig_CAM}, for $\ell=1$, the presence of the outer branch poles with $\text{Re}\, m_n>0$ ( corotating modes) and  $\text{Re}\, m_n <0$ (counterrotating  modes), together with the fundamental poles, contributes to the oscillatory pattern in the angular regions $0<\phi < \pi$  and $\pi<\phi < 2\pi$, respectively.
Similarly, for $\ell=0$,  the outer branch poles are also in charge of the oscillatory pattern.
In addition, in the nonrotating background, contributions from the inner branch poles shift the scattering amplitude to a larger value, whereas in the rotating  background, their effects are negligible.

\begin{figure}[t]
 	\includegraphics[width=\linewidth]{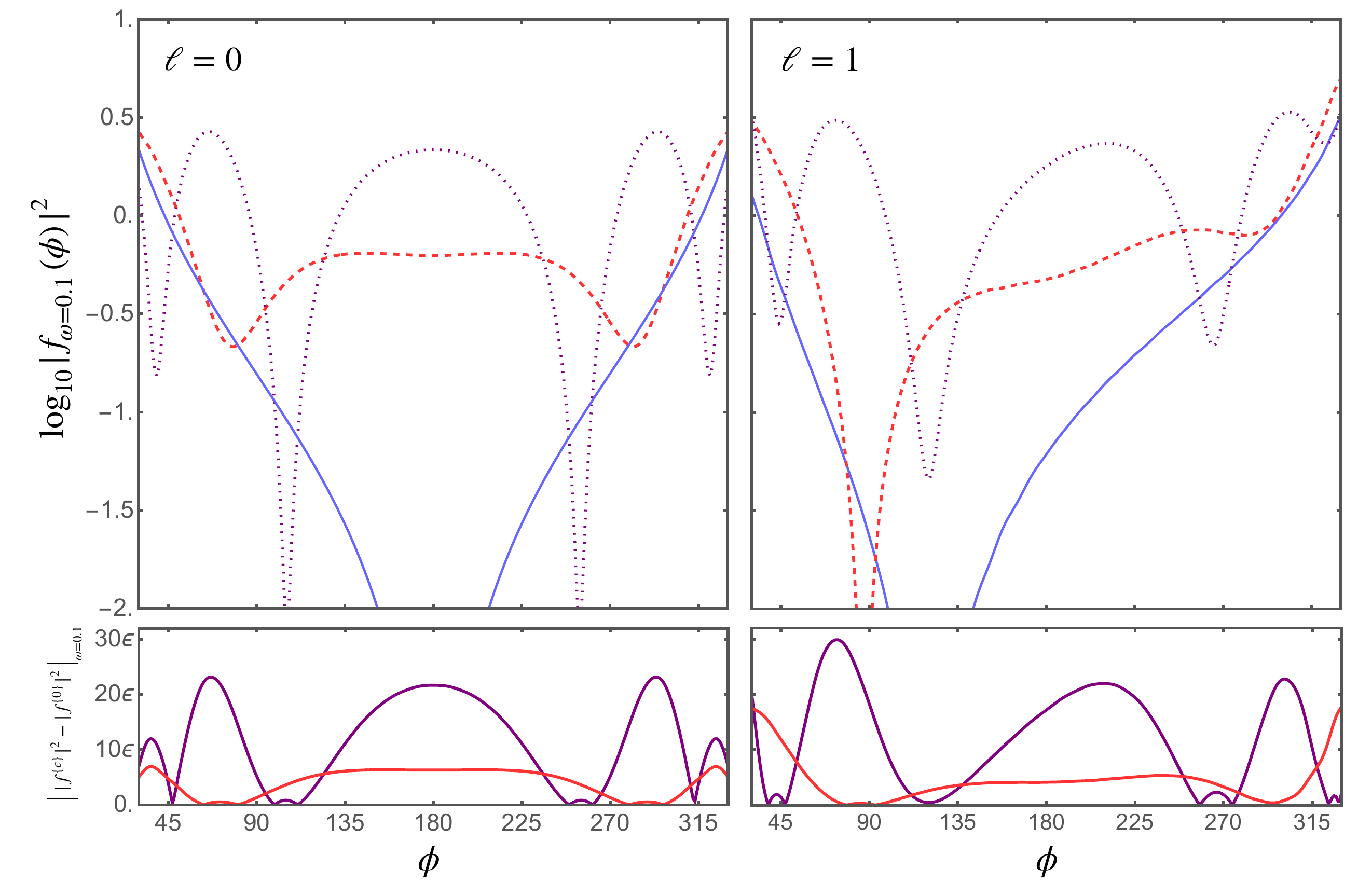}
 	\caption{ The logarithm of the squared scattering amplitude is shown for  low frequency $\omega = 0.1$ in  the nonrotating  (left) and the rotating (right) backgrounds, including both the unperturbed (blue curve) and perturbed cases. The red dashed line corresponds to $r_0 = 10$, while the purple dot-dashed line is for $r_0 = 30$. ($\alpha=3/2$, $\epsilon=10^{-2}$)}
 	\label{fig_obs_lowf}
 \end{figure}

\begin{figure}[t]
	 	\includegraphics[width=0.9\linewidth]{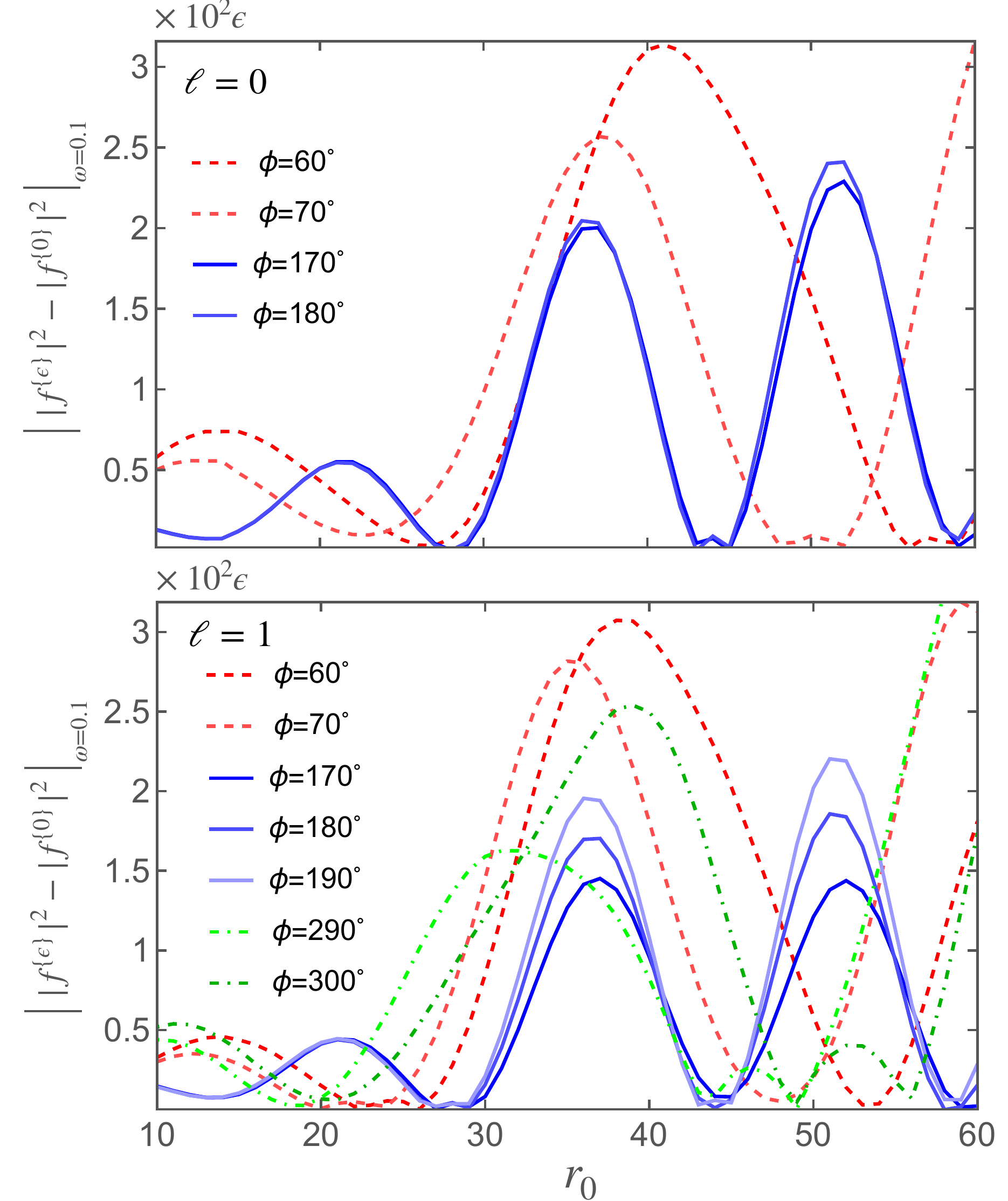}
 	\caption{The deviation of the scattering amplitude due to the perturbations for low frequency $\omega=0.1$  in both the nonrotating  (top) and rotating (bottom) backgrounds at various scattering angles. The parameters are the same as in Fig.~	\ref{fig_obs_lowf}.}
 	\label{fig_obs_angle_lowf}
 \end{figure}

\begin{figure}[t]
	\includegraphics[width=0.9\linewidth]{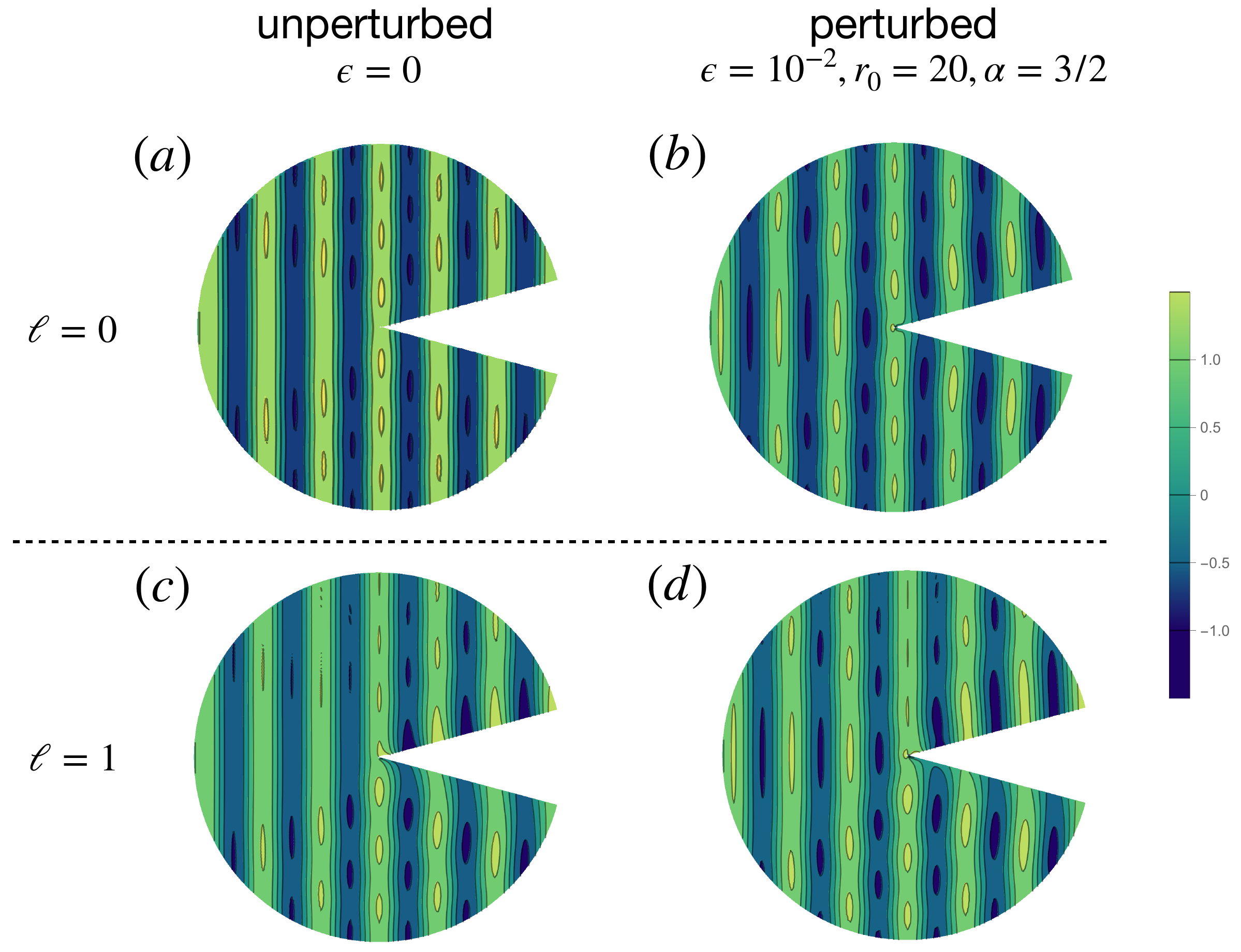}
	\caption{The wave scattering interference patterns are shown for both unperturbed and perturbed cases. Planar waves with frequency $\omega=0.1$ are incident from the left, and scattered by the analog nonrotating [(a),(b)] and rotating [(c),(d)] black holes. In the perturbed case, the magnitude of the large-angle scattering show  the non-negligible  difference from that of the unperturbed case. The blank region in the small angle is intentionally excluded to avoid numerical errors.}
	\label{fig_scattering}
\end{figure}

As for low frequency, the scattering amplitude is shown in Fig.~\ref{fig_obs_lowf}.  For the unperturbed cases in both nonrotating and rotating backgrounds, the scattering amplitudes are consistent with these reported in Ref.~\cite{Dolan2013b}.
 When turning on the space-dependence perturbations, the effects of the perturbations seem to give significant change in the scattering amplitude at all range of the scattering angles in Fig.~\ref{fig_obs_lowf}.
 In principle, overtone destabilization in Fig.~\ref{fig_traj_lowf}  leads to the large perturbed effects on the scattering amplitude at small scattering angles,  whereas the fundamental RP  destabilization in Fig.~\ref{fig_fund}  leave an imprint on the scattering amplitude at large scattering angles, as summarized in Fig.~\ref{fig_obs_angle_lowf}.
Notably, we have found that the fundamental RP is more stable in the corotating mode than in the counterrotating  mode, which is reflected the fact that the perturbed effect on the scattering amplitude is greater for $\phi \lesssim \pi$ compared to $\phi \gtrsim \pi$ ( see Fig.~\ref{fig_obs_angle_lowf}).
The overall effect can be seen in the scattering interference pattern in Fig.~\ref{fig_scattering}, where the perturbations induce a substantial effect at large angles, potentially observable in experiments.


\section{Concluding remarks}\label{secvi}
In this article, we used a simple analog system to study the resonance spectrum in terms of RPs, which could enhance our understanding of resonances in the context of a rotating black hole.  We studied the destabilization of the RP spectrum and the associated observations of the scattering amplitude in the DBT geometry. We use a spatially tunable Rabi coupling in a two-component BEC system to study the gapped excitations of the condensate. We consider the mass shell as a perturbation to mimic the environmental effects, giving a bump potential. We  first compute the scattering amplitude with a homogeneous mass effect using the semiclassical approximation for high-frequency scattering to provide an alternative interpretation of the scattering amplitude in terms of orbiting, revealing clear signatures in the scattering amplitude.

A bifurcation in the RP spectrum occurs in the complex-$m$ plane in the presence of a bump potential, which is an indicator of destabilization of the RP spectrum.
When this occurs, the destabilization criterion, $\vert\Delta \lambda_{\pm,n}^{{\epsilon}}/\lambda_{\pm,n}^{(0)}\vert\gg \epsilon$ holds.
 Using the CAM method, it is demonstrated that the poles of the outer branch, together with the fundamental RP in the bifurcated spectrum, cause additional oscillations in the scattering amplitude. This contrasts with the interpretation derived from the  semiclassical approximation.
We also study the migration of RPs by shifting the bump position.
Our results show that the RPs of the corotating modes are more stable than those of the counterrotating  modes. Large migrations and overtaking jumps of the overtones (fundamental RP) leave an imprint on the scattering amplitude at small (large) scattering angles. This effect can be observed in the scattering interference pattern in experiments.

Future research will investigate the static and dynamic tidal Love numbers in this draining bathtub model under the influence of the environment, as described in Refs.~\cite{Luca2025, Hui2021, Chakraborty2024}. This will help us understand the tidal responses of an analog black hole, providing valuable insights into related astrophysical phenomena.

\section*{acknowledgments}
This work was supported in part by the National Science and Technology Council (NSTC) of Taiwan, Republic of China under the grant No. 114-2112-M-160 -001 -MY2 (W.C.S.) and 114-2112-M-259 -007 -MY3 (D.S.L.).

\appendix

\section{NUMERICAL METHOD FOR COMPUTING
	REGGE POLES}\label{appA}

For completeness, let us review the numerical method.
The approach originated from \cite{Leaver1990}, can be extended  to the case of  a mass shell around the black hole.
We follow the numerical treatments of calculating the QNM and the RP spectrum for a black hole surrounded by thin-shell matter in Refs.~\cite{Torres2023, Torres2023b}
The solution of the Klein-Gordon equation \eqref{KGE} can be written as a series expansion around the point $r=b$ selected to be outside the mass shell where $b>r_0+1/\alpha$ in \eqref{mass}.

\begin{align}
	H_{\omega,m}(r)=e^{i\omega r^\ast(r)}\sum_{k=0}^{\infty}a_k\left(1-\frac{b}{r}\right)^k.
	\label{happ}
\end{align}

After substituting it into the \eqref{KGE}, we have four-term recurrence relation given by
\begin{align}
	\alpha_ka_{k+1}+\beta_ka_k+\gamma_ka_{k-1}+\delta_ka_{k-2}=0,
	\label{re_rel}
\end{align}
for $k=2,3,\cdots$, where
\begin{align}
	\alpha_k=&-\frac{4 \left(b^2-1\right)^2 k (k+1)}{b}\, ,\\
	\beta_k=&\frac{8 \left(b^2-1\right) k \left[-i b^3 \omega +\left(b^2-3\right) k+1\right]}{b},\\
	\gamma_k=&6 b \left[4 k (2 k-3)+5\right]-\frac{5 \left[4 k (3 k-5)+9\right]}{b}\nonumber\\
	&-4 b \left(l^2+1\right) m^2-16 i b^2 (k-1) \omega\nonumber\\
	&+b^3 \left[-4 (k-1) k+4 m (2 l \omega +m)-1\right]\, ,\\
	\delta_k=&\frac{20 (3-2 k)^2}{b}+8 i b^2 (k-2) \omega \nonumber\\
	&+4 b \left[-8 k^2+22 k+2 \left(l^2+1\right) m^2-15\right].
\end{align}
with the initial conditions
\begin{align}
	&a_0=e^{-i\omega r^\ast(b)}H_{\omega,m}(b),\\
	&a_1=be^{-i\omega r^\ast(b)}\left[H'_{\omega,m}(r)-\frac{i\omega}{r^2-1}H_{\omega,m}(r)\right]_{r=b},
\end{align}
which can be found numerically by integrating \eqref{KGE} from the horizon $r=1$ up to $r=b>r_0+1/\alpha$.

Furthermore, one can use Gaussian elimination to reduce \eqref{re_rel} to the three-term recurrence relation
\begin{align}
	&\alpha_0'a_1+\beta_0'a_0=0,\\
	&\alpha_k'a_{k+1}+\beta_k'a_k+\gamma_k'a_{k-1}=0, \quad k=1,2,\cdots.
\end{align}
$\alpha_k',\,\beta_k'$ and $\gamma_k'$ can be expressed in terms of $\alpha_k,\,\beta_k,\,\gamma_k$, and $\delta_k$, given by
\begin{align}
	&\alpha_0=\alpha_0', \,\beta_0'=\beta,\,\gamma_0'=\gamma_0\\
	&\alpha_1=\alpha_1', \,\beta_1'=\beta_1,\,\gamma_1'=\gamma_1\\
	&\alpha_k'=\alpha_k,\,\beta_k'=\beta_k-\alpha_{k-1}'\delta_k/\gamma_{k-1}',\quad{\text{and}}\nonumber\\
	&\gamma_k'=\gamma_k-\beta_{k-1}'\delta_k/\gamma_{k-1}'.
\end{align}

Regarding to the issue of convergence of series expansion (S34), we refer the study of Ref.~\cite{Benhar1999} where mentioned that $a_k$ is a minimal solution to the recurrence relation when $b/2 < r_0+1/\alpha < b$ that in turn gives the continued fraction as
\begin{align}
	\frac{a_1}{a_0}=-\frac{\gamma_1'}{\beta_1'-}\frac{\alpha_1'\gamma_2'}{\beta_2'-}\frac{\alpha_2'\gamma_3'}{\beta_3'-}\cdots.
\end{align}

This modified continued fraction method  can calculate not only quasibound state but also the destabilized QNMs and RPs \cite{Cheung2022}.\\
\begin{widetext}
\section{MORE DETAILED DERIVATION OF  (\ref{lambda})}\label{appB}
Here we provide more detailed derivations to achieve at (\ref{lambda}).  We start from the phase shift $\delta_{\pm m}$ and $\xi_n^{\pm}$,
	\begin{align}
\xi_0^{+} - \xi_1^{-}
=& \omega \bigg\{ \bar{b}_- \left(e^{\frac{\bar{b}_-+\phi -2 \pi }{\bar{a}_-}}+b_{-c}\right)+\phi  \left(e^{\frac{\bar{b}_-+\phi -2 \pi }{\bar{a}_-}}+b_{-c}\right)-2 \pi  \left(e^{\frac{\bar{b}_-+\phi -2 \pi }{\bar{a}_-}}+b_{-c}\right)\nonumber\\
&+\bar{a}_- e^{-\frac{2 \pi }{\bar{a}_-}} \left[e^{\frac{\bar{b}_-+\phi }{\bar{a}_-}}-e^{\frac{\bar{b}_-+\phi }{\bar{a}_-}} \log \left(e^{\frac{\bar{b}_-+\phi -2 \pi }{\bar{a}_-}}\right)+e^{\frac{2 \pi }{\bar{a}_-}} b_{-c}\right]-\bar{b}_+ \left(e^{\frac{\bar{b}_+-\phi }{\bar{a}_+}}+b_{+c}\right)\nonumber\\
&+\phi  \left(e^{\frac{\bar{b}_+-\phi }{\bar{a}_+}}+b_{+c}\right)-\bar{a}_+ \left[e^{\frac{\bar{b}_+-\phi }{\bar{a}_+}}-e^{\frac{\bar{b}_+-\phi }{\bar{a}_+}} \log \left(e^{\frac{\bar{b}_+-\phi }{\bar{a}_+}}\right)+b_{+c}\right] \bigg\}\nonumber\\
&+\omega \left[\frac{\hat{a}_- (\bar{b}_-+\phi -2 \pi )}{\bar{a}_-}+\frac{\hat{a}_+ (\phi -\bar{b}_+)}{\bar{a}_+}\right] \frac{\epsilon}{E^2}  + \mathcal{O}(\epsilon^2).
	\end{align}
Then, the difference can be expanded as the following form
\begin{equation}
\xi_0^{+} - \xi_1^{-}
\approx (A + B \phi) + \frac{\epsilon}{E^2} (A' + B' \phi) +\mathcal{O}(\phi^2),
\end{equation}
%where the nonperturbative term with $\bar{a}_s>0$ and $\bar{b}_s<0$, (note that, the scale is about $b_{sc}=2$, $e^{\frac{\bar{b}_+}{\bar{a}_+}} = 0.752767$, and $e^{\frac{\bar{b}_--2 \pi }{\bar{a}_-}} = 0.000104141$ for $C=0$; $b_{+c}=0.828427$, $b_{-c}=4.82843$, $e^{\frac{\bar{b}_+}{\bar{a}_+}} = 0.690056$, and $e^{\frac{\bar{b}_--2 \pi }{\bar{a}_-}} = 0.0000109822$ for $C=1$.)
where the unperturbed terms with
\begin{align}
A =& \omega \left[ \bar{a}_- \left(e^{\frac{\bar{b}_- -2 \pi }{\bar{a}_-}} +b_{-c}\right) +\bar{a}_+ \left( -e^{\frac{\bar{b}_+}{\bar{a}_+}} - b_{+c} \right)+(\bar{b}_- -2\pi ) b_{-c}-\bar{b}_+ b_{+c} \right] \nonumber\\
\approx&\, \omega (-2\pi b_{-c})\nonumber \\
\approx&-4 \omega r_e \pi \, ,
\end{align}
\begin{equation}
\begin{split}
B &= \omega \left(b_{+c} + b_{-c} + e^{\frac{\bar{b}_--2 \pi }{\bar{a}_-}} + e^{\frac{\bar{b}_+}{\bar{a}_+}} \right) \\
&\approx \omega (b_{+c} + b_{-c}) \\
&= 4 \omega \sqrt{D^2+C^2} \\
&= 4 \omega r_e \,,
\end{split}
\end{equation}
and the perturbed terms with
\begin{equation}
\begin{split}
A' &= \omega \left[\frac{\hat{a}_- (\bar{b}_--2 \pi )}{\bar{a}_-}-\frac{\hat{a}_+ \bar{b}_+}{\bar{a}_+} \right] ,\\
B' &= \omega \left[\frac{\hat{a}_-}{\bar{a}_-}+\frac{\hat{a}_+}{\bar{a}_+} \right] \, .
\end{split}
\end{equation}
Therefore, the wavelength is given by $B+\frac{\epsilon}{E^2} B'=2\pi/\lambda$,
\begin{equation}
\begin{split}
\lambda \approx \frac{\pi}{2\omega r_e} -\frac{\epsilon}{E^2}\, \left(\frac{\hat{a}_-}{\bar{a}_-}+\frac{\hat{a}_+}{\bar{a}_+} \right) \frac{\pi}{8 \omega^2 r_e^2}
\end{split}
\end{equation}
with the small mass correction.\end{widetext}

\bibliography{ABH_RPs.bib}

\end{document}